\documentclass[aps,prd,groupedaddress,showpacs,showkeys]{revtex4}
\usepackage{amssymb}
\usepackage{epsfig}

\newcommand{\eq}{\begin{eqnarray}}
\newcommand{\en}{\end{eqnarray}}

\newcommand{\la}{\langle} 
\newcommand{\ra}{\rangle} 
\newcommand{\beqn}{\begin{eqnarray}}
\newcommand{\eeqn}{\end{eqnarray}}
\newcommand{\beq}{\begin{equation}}
\newcommand{\eeq}{\end{equation}}
\newcommand{\barr}{\begin{array}}
\newcommand{\earr}{\end{array}}

\begin{document}

\title{Magnetic moments of heavy baryons  
in the relativistic three-quark model}

\vspace*{1.2\baselineskip}

\author{Amand Faessler$^1$, Th. Gutsche$^1$, 
M. A. Ivanov$^2$, J. G. K\"{o}rner$^3$, 
V. E. Lyubovitskij$^1$\footnote{On leave of absence  
from Department of Physics, Tomsk State University, 
634050 Tomsk, Russia}, 
D. Nicmorus$^1$\footnote{On leave of absence 
from Institute of Space Sciences, P.O. Box MG-23, 
Bucharest-Magurele 76900 Romania}, 
K. Pumsa-ard$^1$ 
\vspace*{1.2\baselineskip}}

\affiliation{$^1$ Institut f\"ur Theoretische Physik,
Universit\"at T\"ubingen,
\\ Auf der Morgenstelle 14, D-72076 T\"ubingen, Germany 
\vspace*{1.2\baselineskip} \\
$^2$ Bogoliubov Laboratory of Theoretical Physics,
Joint Institute for Nuclear Research,~141980~Dubna,~Russia 
\vspace*{1.2\baselineskip} \\
$^3$ Institut f\"{u}r Physik, Johannes Gutenberg-Universit\"{a}t,
D-55099 Mainz, Germany\\}

\date{\today}

\begin{abstract}
The magnetic moments of ground state single, double and triple 
heavy baryons containing charm or bottom quarks 
are calculated in a relativistic three-quark model, 
which, in the heavy quark limit, is consistent with 
Heavy Quark Effective Theory and Heavy Hadron Chiral 
Perturbation Theory.
The internal quark structure of baryons is modeled 
by baryonic three-quark currents with a spin-flavor structure
patterned according to standard covariant baryonic wave functions 
and currents used in QCD sum rule calculations.
\end{abstract}


\maketitle

\section{Introduction}

Electromagnetic properties of baryons are an important source 
of information on their internal structure. The success of 
quark models for the description of static properties 
(masses, magnetic moments, etc.) and the results from
deep inelastic lepton scattering are clear indication 
for the three-quark structure of baryons. 
 
Magnetic moments of heavy baryons (mostly with a single 
heavy quark) have been considered in different approaches. 
In Refs.~\cite{Choudhury:1976dn}-\cite{Verma:1981sx} 
the magnetic moments of charmed baryons have been computed 
using naive quark models based on different 
realizations of spin-flavor symmetry. 
In Refs.~\cite{Bose:1980vy}-\cite{Julia-Diaz:2004vh} 
the magnetic moments of charmed and bottom hyperons have been 
calculated in quark models incorporating the ideas of 
hadronization, confinement, chiral symmetry and Poincar\'{e} 
covariance. In Refs.~\cite{Oh:1991ws}-\cite{Scholl:2003ip} 
soliton-type approaches were applied in the analysis 
of the magnetic moments of heavy baryons. 
In Refs.~\cite{Zhu:1997as} QCD spectral sum rules in the 
presence of the external electromagnetic field have been used 
to calculate the magnetic moments of the $\Sigma_c$ and $\Lambda_c$ 
baryons. In~\cite{Aliev:2001ig} the method of light-cone 
QCD sum rules has been used to calculate the magnetic moments of 
the $\Lambda_{b}$ and $\Lambda_c$ baryons.  

Heavy hadron chiral perturbation theory (HHChPT)~\cite{Wise:1992hn} 
has been applied in Refs.~\cite{Cho:1992nt}-\cite{Tiburzi:2004mv} 
to derive model-independent expressions for the magnetic moments 
of heavy baryons containing a single heavy quark. 
In HHChPT~\cite{Wise:1992hn}-\cite{Tiburzi:2004mv} 
heavy quark symmetry (HQS) and chiral symmetry ($\chi$S) have been combined 
in order to describe the soft hadronic interactions of hadrons 
containing a heavy quark. The underlying Lagrangian deals with heavy 
hadrons, light mesons and external fields (photons). 

In the heavy quark limit (HQL), when the heavy quark mass goes to infinity 
$(m_Q \to \infty)$, baryons containing a single heavy quark 
can be classified according to the spin of the light degrees of 
freedom: 
i) the antitriplet $\bf \bar{3}$ of baryons for $\Lambda$-type baryons 
(isosinglet $\Lambda_Q$ with quark content Q[ud], and the isodoublet $\Xi_Q$ 
with quark content $Q[us]$ and $Q[ds]$ and ii) the sextet $\bf 6$ of baryons 
or $\Sigma$-type baryons (isotriplet triplet $\Sigma_Q$ with quark 
content $Q\{uu\}$, $Q\{ud\}$ and $Q\{dd\}$, and the isodoublet $\Xi_Q^\prime$ 
with quark content $Q\{us\}$ and $Q\{ds\}$). Here the symbols 
$[ \ldots ]$ and  $\{ \ldots \}$ denote antisymmetric and symmetric 
flavor index combinations. For the $\bf \bar{3}$ states the total spin 
of the light diquark system is $s_l = 0$, while for the $\bf 6$ states 
the total spin of the light diquark system is $s_l = 1$. 
In the case of the $\bf \bar{3}$ 
states, the spin of the baryon is made up entirely by the heavy quark and, 
therefore, the magnetic moment of these baryons vanishes in the 
HQL~~\cite{Cho:1992nt,Savage:1994zw} (the magnetic moment is a spin--flip
transition down by a factor of $1/m_Q$). 
The leading $O(1/[m_Q \Lambda_\chi])$ long-distance 
contribution to the magnetic moments of $\Lambda$-type baryons arises from 
spin-symmetry breaking which also leads to the $\Sigma_Q^\ast - \Sigma_Q$ mass 
splitting~\cite{Savage:1994zw}, where 
$\Lambda_{\chi} = 4 \pi F_\pi \simeq 1.2$ GeV is the scale 
parameter of the spontaneously broken chiral symmetry and $F_\pi$ is the 
pion decay constant. In the case of the $\Sigma$-type baryons 
the magnetic moments are nonzero in the HQL due to the contribution of 
the light-quark system~\cite{Cho:1992nt}. In particular, the leading 
contributions from the light- and heavy-quark magnetic interactions are 
of order of $O(1/\Lambda_\chi)$ and 
$O(1/m_Q)$~\cite{Cho:1992nt}, respectively. 
The next-to-leading chiral corrections to the magnetic moments of 
$\Sigma$-type and $\Lambda$-type baryons are of order 
$O(1/\Lambda_\chi^2)$ and $O(1/[m_Q \Lambda_\chi^2])$, 
respectively, the contributions of which have been calculated in 
Ref.~\cite{Banuls:1999mu}. 

The present paper focuses on the magnetic moments of the ground 
state single, double and triple heavy baryons in the relativistic 
three-quark model (RQM) which has been developed both for
light~\cite{Ivanov:1996pz} and heavy 
baryons~\cite{Ivanov:1996fj}-\cite{Faessler:2001mr}. 
This model can be viewed as an quantum 
field theory approach based on an interaction Lagrangian of light and 
heavy baryons interacting with their constituent quarks. 
The coupling strength of the baryons with
the three constituent quarks is determined by the 
compositeness condition $Z_H=0$, where $Z_H$ is the wave 
function renormalization constant of the hadron. The condition $Z_H=0$ has  
been proposed in~\cite{Weinberg:1962hj} and extensively used 
in~\cite{Efimov:1993ei}. 
The compositeness condition enables one to unambiguously and 
consistently relate approaches with quark and hadron degrees of 
freedom to the effective Lagrangian approaches formulated in 
terms of hadron variables only. 
Our strategy is as follows. We start with an interaction
Lagrangian written down in terms of quark and hadron variables.
Then, by using Feynman rules, the $S$-matrix elements describing
hadron-hadron interactions are given in terms of a set of quark 
diagrams. The compositeness condition ensures that there is no 
double counting of the quark and hadron degrees of freedom.
One of the corollaries of the compositeness condition is the absence
of a direct interaction of the dressed charged particle with the
electromagnetic field. 
The RQM model contains only a few model parameters: the masses of 
the light and heavy quarks and certain size parameters that define 
the size of the distribution of the constituent quarks inside the 
hadron. The RQM approach has been previously used to
compute the exclusive semileptonic, nonleptonic, strong and 
electromagnetic decays of single heavy 
baryons~\cite{Ivanov:1996fj}-\cite{Ivanov:1999pz} in the heavy
quark limit  $m_Q\to\infty$ always employing the same set of 
model parameters. In Ref.~\cite{Ivanov:1999bu} the approach has 
been extended to the study of heavy baryon transitions at finite 
values of the heavy quark mass without using an explicit $1/m_Q$ 
expansion. Later, in Ref.~\cite{Faessler:2001mr} the RQM has been 
generalized to the study of double heavy baryons. In the present 
manuscript we extend our approach to the triple heavy baryons. 
We also check the consistency of our approach with HHChPT.  

We proceed as follows. First, in Section II  we briefly explain 
the basic ideas of the RQM. Next, in Section III we discuss 
matrix elements for the baryon mass operator and the electromagnetic 
vertex function. In Section IV we discuss the matching of 
our model to Heavy Hadron Chiral Perturbation Theory (HHChPT). 
In Section IV we compare our numerical results with the 
results of other theoretical approaches. 
Finally, we summarize our results in Section V. 
 
\section{Relativistic Three-Quark Model}

We will consistently employ the relativistic three-quark model 
(RQM)~\cite{Ivanov:1996fj}-\cite{Faessler:2001mr} to compute 
the magnetic moments of single, double and triple heavy baryons. 
In the following we will present details of the model which is
essentially based on an interaction Lagrangian describing
the coupling between baryons and their constituent quarks.

The coupling of a baryon $B(q_1q_2q_3)$ to its constituent 
quarks $q_1$, $q_2 $ and $q_3$ is described by the Lagrangian 
\eq\label{Lagr_str}
{\cal L}_{\rm int}^{\rm str}(x) = g_B \bar B(x) \, 
\int\!\! dx_1 \!\! \int\!\! dx_2 \!\! \int\!\! dx_3 \, 
F_B(x,x_1,x_2,x_3) \, J_B(x_1,x_2,x_3) \, + \, {\rm H.c.}  
\en 
where $J_{B}(x_1,x_2,x_3)$ is the three-quark current with the 
quantum numbers of the relevant baryon $B$. One has
\eq 
J_{B}(x_1,x_2,x_3) \, = \, \epsilon^{a_1a_2a_3} \, 
\Gamma_1 \, q^{a_1}_1(x_1) \, q^{a_2}_2(x_2) 
C \, \Gamma_2 \, q^{a_3}_3(x_3) \, ,
\en 
where $\Gamma_{1,2}$ are Dirac structures, $C=\gamma^{0}\gamma^{2}$ is
the charge conjugation matrix and $a_i, i=1,2,3$ are color indices.

The function $F_B$ is related to the scalar part of the 
Bethe-Salpeter amplitude and characterizes the finite size 
of the baryon. To satisfy translational invariance the function 
$F_B$ has to fulfill the identity 
\eq\label{trans_inv}
F_B(x+a,x_1+a,x_2+a,x_3+a) \, = \, 
F_B(x,x_1,x_2,x_3) 
\en
for any 4-vector $a\,$. 
In the following we use a particular form for the vertex function 
\eq\label{vertex}
F_B(x,x_1,x_2,x_3) \, = \, \delta^4(x - \sum\limits_{i=1}^3 w_i x_i) \;  
\Phi_B\biggl(\sum_{i<j}( x_i - x_j )^2 \biggr) 
\en 
where $\Phi_B$ is the correlation function of three constituent 
quarks with masses $m_1$, $m_2$, $m_3$. The variable $w_i$ is defined by 
$w_i=m_i/(m_1+m_2+m_3)$ 
and therefore depends only on the relative Jacobi coordinates $(\xi_{1},
\xi_{2})$ as $\Phi_{B}(\xi^2_1+\xi^2_2)$, where 
\eq\label{coordinates}
x_1&=&x \, - \, \frac{\xi_1}{\sqrt{2}} \, (w_2+w_3) 
        \, + \, \frac{\xi_2}{\sqrt{6}} \, (w_2 - w_3)\,,\nonumber\\
x_2&=&x \, + \, \frac{\xi_1}{\sqrt{2}} \,  w_1      
        \, - \, \frac{\xi_2}{\sqrt{6}} \, (w_1 + 2w_3)\,, \\
x_3&=&x \, + \, \frac{\xi_1}{\sqrt{2}} \,  w_1    
        \, + \, \frac{\xi_2}{\sqrt{6}} \, (w_1 + 2w_2)\,,\nonumber
\en 
and where $x \,= \, \sum\limits_{i=1}^3 w_i x_i$ is 
the center of mass (CM) coordinate. In terms of the CM and quark 
coordinates the Jacobi coordinates are simply given by 
\eq\label{Jacobi_cs} 
\xi_1 = \frac{1}{\sqrt{2}} \, ( x_2 + x_3 - 2 x_1 )\,, \quad\quad 
\xi_2 = \sqrt{\frac{3}{2}} \, ( x_3 - x_2 ) \,. 
\en 
Note, that the choice of Jacobi coordinates is not unique. 
We choose the most convenient ansatz defined by 
Eqs.~(\ref{coordinates}) and (\ref{Jacobi_cs}). 
Expressed in relative Jacobi coordinates and the center of mass 
coordinate, the Fourier transform of the vertex function 
reads~\cite{Ivanov:1996fj}-\cite{Faessler:2001mr}:  
\eq 
\Phi_{B}(\xi^2_1+\xi^2_2)=
\int\!\frac{d^4p_1}{(2\pi)^4}\!\int\!\frac{d^4p_2}{(2\pi)^4}
e^{-ip_1\xi_1-ip_2\xi_2}{\widetilde{\Phi}}_{B}(-p^2_1-p^2_2)
\en 
In the numerical calculations we consider two specific limits 
for the correlation functions (CF) of heavy baryons: 
i) the exact form with no approximations (full) and ii)
the heavy quark limit (HQL) 
for the heavy 
quark masses in the baryonic correlation functions (BCF). We will 
refer to the second model as the HQL BCF model.  
In particular, the HQL BCF model for single heavy 
baryons means that: 1) they are treated as bound states of a heavy quark 
and the light diquark system; 2) a heavy quark is located in the center 
of the system  and is surrounded by the light degrees of freedom. 
We apply the heavy quark limit $m_1 = m_Q \to \infty$ 
in Eq.~(\ref{coordinates}), i.e. $w_1 \to 1$, $w_2 \to 0$ 
and $w_3 \to 0$ and, therefore, one has $x_1 \to x$. 
Double heavy baryons are treated as bound states of a heavy diquark 
and a light quark. In the HQL BCF model the heavy diquark is located at the 
center of 
the double heavy baryon:  $m_2 = m_Q \to \infty$, 
$m_3 = m_{Q^\prime} \to \infty$ and, therefore, one has 
$w_1 \to 0$, $w_2 \to 1/2$ and $w_3 \to 1/2$. 
In the HQL BCF model the triple heavy baryon $\Omega_{ccb}^+$ is treated as a 
bound 
state of a heavy bottom quark and a relatively ``light'' charm diquark 
with $m_1 = m_b \to \infty$ and, therefore, $w_1 \to 1$, $w_2 \to 0$, 
$w_3 \to 0$ and  $x_1 \to x$ (its structure is similar 
to the single heavy baryon in HQL). Finally, the HQL BCF limit for the 
$\Omega_{cbb}^0$ baryon means that the $\Omega_{cbb}^0$ is a bound state of 
a heavy bottom diquark and and a relatively ``light'' charm quark with 
$m_2 = m_3 = m_b \to \infty$ and, therefore, $w_1 \to 0$, $w_2 \to 1/2$ 
and $w_3 \to 1/2$ (its structure is similar 
to the double heavy baryon in the HQL). 

We consider two types of heavy baryons: 
$\Lambda$-type baryons which are bound states of a quark and diquark 
system with spin 0 and $\Sigma$-type baryons which are bound states 
of a quark and a diquark system carrying spin 1. In general, for the 
$\Lambda$-type baryons one can construct three types of currents 
without derivatives - pseudoscalar $J^P$, scalar $J^S$ and 
axial-vector $J^A$ and two types of currents (the vector $J^V$ and 
the tensor $J^T$ form ) for $\Sigma$-type baryons 
(see Refs.~\cite{Shuryak:1980pg}-\cite{Groote:1996xb}  
and \cite{Ivanov:1996fj}-\cite{Faessler:2001mr}):   
\eq 
J^P_{\Lambda_{q_{1}[ q_{2}q_{3}] }}&=& \epsilon^{a_1a_2a_3} \, 
q_{1}^{a_1} \, q_{2}^{a_2} C \gamma_5 q_{3}^{a_3} \,, 
\nonumber\\
J^S_{\Lambda_{q_{1}[ q_{2}q_{3}] }}&=& \epsilon^{a_1a_2a_3} \, \gamma^5 \, 
q_{1}^{a_1} \, q_{2}^{a_2} C  q_{3}^{a_3} \,,\\
J^A_{\Lambda_{q_{1}[ q_{2}q_{3}] }}&=& \epsilon^{a_1a_2a_3} \, 
\gamma^\mu \, q_{1}^{a_1} \, q_{2}^{a_2}  
C\gamma_\mu\gamma_5 q_{3}^{a_3} \,, 
\nonumber
\en 
and 
\eq 
J^V_{\Sigma_{q_{1}\{ q_{2}q_{3}\} }}&=&\varepsilon^{a_1a_2a_3} 
\gamma^\mu\gamma^5 q_{1}^{a_1} q_{2}^{a_2} C \gamma_\mu q_{3}^{a_3} \,, 
\nonumber\\
J^{T}_{\Sigma_{q_{1}\{ q_{2}q_{3}\} }}&=&\varepsilon^{a_1a_2a_3} 
\sigma^{\mu\nu}\gamma^5
q_{1}^{a_1} q_{2}^{a_2} C \sigma_{\mu\nu} q_{3}^{a_3}  \,. 
\en 
The symbols $[...]$ and $\{...\}$ denote antisymmetrization
and symmetrization over flavor indices of the second and the
third quark, respectively. 

In the following we restrict ourselves 
to the simplest baryonic currents - 
pseudoscalar $J^P$ for the $\Lambda$-type and 
vector $J^V$ for the $\Sigma$-type baryons. Note, that these 
currents for the $\Lambda$ and $\Sigma$-type baryons have the 
correct nonrelativistic limit (NL). In particular, in the case 
of the $\Lambda$-type baryons the scalar current goes to zero 
in the nonrelativistic limit whereas the pseudoscalar and axial-vector 
currents become degenerate in this limit with the following 
naive quark model baryon spin-flavor function: 
\eq\label{LambdaQ_wf} 
|\Lambda_{q_1[q_2q_3]} \ra \, = \, 
\frac{1}{2} \, | q_1 ( q_2 q_3 - q_3 q_2 ) \ra \,\,\, 
| \uparrow (\uparrow \downarrow - \downarrow \uparrow) \ra \,. 
\en 
In the case of the $\Sigma$-type baryons the vector and tensor currents 
also become degenerate in the nonrelativistic limit. Their 
naive quark model spin-flavor wave functions read:
\eq\label{SigmaQ_wf}
|\Sigma_{q_1\{q_2q_3\}} \ra \, = \, \frac{1}{2\sqrt{3}} \, 
| q_1 ( q_2 q_3 + q_3 q_2 ) \ra \,\,\, 
| \uparrow (\uparrow \downarrow + \downarrow \uparrow)  
- 2 \downarrow \uparrow \uparrow \ra \,. 
\en 
The classification of the heavy baryon states (spin-parity, flavor content and 
quantum numbers, mass spectrum) is given in  Table 1 (single 
charm baryons), Table 2 (single bottom baryons) and Table 3 (double 
and triple heavy baryons). We use the data from 
Refs~\cite{Eidelman:2004wy}-\cite{Kiselev:2001fw}  
and restrict ourselves to $\frac{1}{2}^+$ baryons. 

The heavy-baryon quark coupling constants $g_B$ are determined by 
the compositeness condition~\cite{Ivanov:1996fj}-\cite{Faessler:2001mr}
(see also~\cite{Weinberg:1962hj,Efimov:1993ei}). 
The compositeness
condition implies that the renormalization constant of the hadron wave
function is set equal to zero: 
\eq 
Z_B = 1 - \Sigma^\prime_B(m_B) = 1 - g_B^2 \Pi^\prime_B(m_B) = 0 \, 
\en 
where $\Sigma^\prime_B$ is the derivative of the baryon mass 
operator described by the diagram Fig.1 and $m_B$ is the 
heavy baryon mass. 
To clarify the physical meaning of this condition, we first want to remind 
the reader that the renormalization constant $Z_B^{1/2}$ can also be 
interpreted as  
the matrix element between the physical and the corresponding bare state.  
For $Z_B=0$ it then follows that the physical state does not contain  
the bare one and is described as a bound state. 
The interaction Lagrangian Eq.~(\ref{Lagr_str}) and 
the corresponding free parts describe  
both the constituents (quarks) and the physical particles (hadrons)
which are taken to be the bound states of the constituents.
As a result of the interaction, the physical particle is dressed, 
i.e. its mass and its wave function have to be renormalized. 
The condition $Z_B=0$ also effectively excludes
the constituent degrees of freedom from the physical space
and thereby guarantees that there is no double counting
for the physical observable under consideration.
In this picture the constituents exist in virtual states only. 
One of the corollaries of the compositeness condition is the absence 
of a direct interaction of the dressed charged particle with the 
electromagnetic field. Taking into account both the tree-level
diagram and the diagrams with the self-energy insertions into the 
external legs (that is the tree-level diagram times $(Z_B -1$)) one obtains 
a common factor $Z_B$  which is equal to zero. 

We use the standard free fermion Lagrangian for the 
baryons and quark fields: 
\eq
{\cal L}_{\rm free}(x) =  
\bar B(x) (i \not\!\partial - m_B) B(x) + 
\sum\limits_q \, \bar q(x) (i \not\!\partial - m_q) q(x) \, , 
\en 
where $m_q$ is the constituent quark mass. This leads to 
the free fermion propagator for the 
constituent quark: 
\eq\label{quark_propagator} 
i \, S_q(x-y) = \langle 0 | T \, q(x) \, \bar q(y)  | 0 \rangle 
\ = \ \int\frac{d^4k}{(2\pi)^4i} \, e^{-ik(x-y)} \ \tilde S_q(k)  
\en 
where 
\eq
\tilde S_q(k) = \frac{1}{m_q-\not\! k -i\epsilon}
\en 
is the usual free fermion propagator in momentum space. 
We shall avoid the appearance of unphysical imaginary parts 
in Feynman diagrams by postulating the condition that 
the baryon mass must be less than the sum of the constituent 
quark masses $M_{B} < \sum_{i}m_{q_{i}}$. As was mentioned before,  
we treat heavy quark masses 
in the baryonic correlation functions in two specific limits: 
i) using their finite values and ii) applying the heavy quark limit (HQL). 
In the case of single heavy baryons 
we also consider the heavy quark limit for the heavy quark propagator. 
Therefore, in the case of single heavy baryons we consider three 
models: i) exact calculations (full); ii) exact form of the heavy 
quark propagator (with a finite value for the heavy quark mass) 
and heavy quark limit for the baryonic correlation function 
(HQL BCF); 3) full heavy quark limit for both the heavy quark propagator 
and the baryonic correlation function (HQL BCF+HQP). 

The interaction with the electromagnetic field is introduced
in two ways. The free Lagrangians of quarks and hadrons
are gauged in the standard manner by using minimal
substitution:
\eq 
\label{photon} 
\partial^\mu B \to (\partial^\mu - ie_B A^\mu) B\,, 
\hspace*{.5cm} 
\partial^\mu \bar B\to (\partial^\mu +ie_B A^\mu) \bar B\,,
\hspace*{.5cm} 
\partial^\mu q_i \to (\partial^\mu - ie_{q_i} A^{\mu}) q_i\,, 
\hspace*{.5cm} 
\partial^\mu \bar q_i \to (\partial^\mu + ie_{q_i} A^{\mu}) q_i\,, 
\en 
where $e_B$ is the electric charge of the baryon $B$ and 
$e_{q_i}$ is the electric charge of the quark with flavor 
$q_i$. 
The interaction of the baryon and quark fields is then specified by 
minimal substitution. The interaction Lagrangian reads
\eq 
\label{L_em_1}
{\cal L}^{\rm em (1)}_{\rm int}(x) = 
e_B \bar B(x) \!\not\!\! A \, B(x)  +  
\sum\limits_q \, e_q \, \bar q(x) \!\not\!\! A  \, q(x) \,. 
\en 
We remind the reader that the electromagnetic field does not directly
couple to the baryon fields in the relativistic three quark model as
explicitly shown in the following.
The gauging of the nonlocal Lagrangian Eq.~(\ref{Lagr_str})
proceeds in a way suggested and extensively used 
in Refs.~~\cite{Ivanov:1996pz,Mandelstam:1962mi,Terning:1991yt}. 
In order to guarantee local invariance of the strong interaction 
Lagrangian one multiplies 
each quark field $q(x_i)$ in ${\cal L}_{\rm int}^{\rm str}$ with a 
gauge field exponentional.
One then has
\eq\label{gauging}
{\cal L}_{\rm int}^{\rm str + em(2)}(x) &=& g_B \bar B(x) \, 
\int\!\! dx_1 \!\! \int\!\! dx_2 \!\! \int\!\! dx_3 \, 
F_B(x,x_1,x_2,x_3) \, \epsilon^{a_1a_2a_3} \, \Gamma_1 
e^{-ie_{q_1} I(x_1,x,P)} \, q^{a_1}_1(x_1) \\ 
&\times& \, e^{-ie_{q_2} I(x_2,x,P)} \, q^{a_2}_2(x_2)  
C \, \Gamma_2 \, 
\, e^{-ie_{q_3} I(x_3,x,P)} \, q^{a_3}_3(x_3)
\, + \, {\rm H.c.}  \nonumber
\en 
where
\eq\label{path}
I(x_i,x,P) = \int\limits_x^{x_i} dz_\mu A^\mu(z). 
\en
Finally, the Lagrangian suitable for the calculation of the 
electromagnetic properties of heavy baryons is given by: 
\eq
{\cal L}_{\rm full}(x) \, = \, {\cal L}_{\rm free}(x) \, + \, 
{\cal L}_{\rm int}^{\rm em(1)}(x)  \, + \, 
{\cal L}_{\rm int}^{\rm str + em(2)}(x) \,. 
\en

It is readily seen that the full Lagrangian is invariant 
under the transformations
\eq 
& &A^\mu(x) \, \to \, A^\mu(x)+\partial^\mu f(x)\,, \hspace*{1cm}
     q_i(x) \, \to \, e^{ie_{q_i} f(x)} q_i(x)\,,   \hspace*{1cm}
\bar q_i(x) \, \to \, \bar q_i(x) \, e^{-ie_{q_i} f(x)}\,, \hspace*{1cm} \\
& &    B(x) \, \to \, e^{ie_B f(x)} \, B(x)\,, \hspace*{1.5cm} 
  \bar B(x) \, \to \, \bar B(x) \, e^{-ie_B f(x)}\,, \nonumber 
\en
where $e_B = \sum\limits_{i=1}^3 e_{q_i}$. 

When expanding the gauge exponential
up to a certain power of $A_\mu$ relevant to the desired order of perturbation 
theory in the given process the second term of the electromagnetic 
interaction Lagrangian
${\cal L}^{em(2)}_{\rm int}$ arises. 
At first sight it appears that the results will depend on the path $P$
taken to connect the end-points in the path integral in 
Eq.~(\ref{path}).  
However, one needs to know only the derivatives of the path integral
expressions when calculating the perturbative series.
Therefore, we use the 
formalism suggested in~\cite{Ivanov:1996pz,Mandelstam:1962mi,Terning:1991yt} 
which is based on the path-independent definition of the derivative of 
$I(x,y,P)$: 
\eq\label{path1}
\lim\limits_{dx^\mu \to 0} dx^\mu 
\frac{\partial}{\partial x^\mu} I(x,y,P) \, = \, 
\lim\limits_{dx^\mu \to 0} [ I(x + dx,y,P^\prime) - I(x,y,P) ]
\en 
where the path $P^\prime$ is obtained from $P$ by shifting the end-point $x$
by $dx$.
The definition (\ref{path1}) 
leads to the key rule
\eq\label{path2}
\frac{\partial}{\partial x^\mu} I(x,y,P) = A_\mu(x)
\en 
which in turn states that the derivative of the path integral $I(x,y,P)$ does 
not depend on the path P originally used in the definition. The non-minimal 
substitution (\ref{gauging}) is therefore completely equivalent to the 
minimal prescription as is evident from the identities (\ref{path1})  
or (\ref{path2}). In Appendix A we demonstrate explicitly how to 
derive the Feynman rules for a non-local coupling of hadrons to photons 
and quarks considering the case of the coupling of a baryon to three quarks 
and a single photon (see Fig.2).  

In the next step we have to specify the vertex function $\tilde\Phi_B$, which 
characterizes the finite size of the baryons.  
In principle, its functional form can be calculated from the solutions
of the Bethe-Salpeter equation for the baryon bound 
states~\cite{Ivanov:1998ya,Alkofer:2004yf}. 
In Refs.~\cite{Anikin:1995cf} it was found that, using various
forms for the vertex function, the basic hadron observables
are insensitive to the details of
the functional form of the hadron-quark vertex form factor.
We will use this observation as a guiding principle and choose a simple
Gaussian form for the vertex function $\tilde\Phi_B$. 
Any choice for $\tilde\Phi_B$ is appropriate  
as long as it falls off sufficiently fast in the ultraviolet region of 
Euclidean space to render the Feynman diagrams ultraviolet finite. 
We employ a Gaussian form for the vertex function 
\eq\label{Gauss_CF}
\tilde\Phi_B(k_{1E}^2,k_{2E}^2  ) 
\doteq \exp( - [k_{1E}^2 + k_{2E}^2]/\Lambda^2_B )\,, 
\en  
where $k_{1E}$ and $k_{2E}$ are the Euclidean momenta. 
Here $\Lambda_{B}$ is a size parameter parametrizing the distribution 
of quarks inside a given baryon. In fact we shall use only a reduced set
of size parameters, namely $\Lambda_{B_{qqq}}$ for the light baryons, 
$\Lambda_{B_{Qqq}}$ for single heavy baryons, 
$\Lambda_{B_{qQQ}}$ for double heavy baryons and 
$\Lambda_{B_{QQQ}}$ for triple heavy baryons $\Lambda_{B_{QQQ}}$.
We use the following set of parameters:  
\begin{equation}
\begin{array}{ccccc}
m_{u(d)} & m_s & m_c & m_b & \\  \hline
$\ \ 0.42\ \ $ & $\ \ 0.57\ \ $ & $\ \ 1.7\ \ $ & $\ \ 5.2\ \ $
&
$\ \ {\rm GeV} $\\
\end{array}\label{fitmas}
\end{equation}
and
\begin{equation}
\begin{array}{ccccc}
\Lambda_{B_{qqq}} & \Lambda_{B_{Qqq}} & 
\Lambda_{B_{qQQ}} & \Lambda_{B_{QQQ}} & \\  
\hline
$\ \ 1.25 \ \ $  & $\ \ 1.8\ \ $    & 
$\ \ 2.5  \ \ $  & $\ \ 5  \ \ $    &$\ \ {\rm GeV} $\\
\end{array}\label{fitlambda}
\end{equation}
The size parameters $\Lambda_{B_{qqq}}$, $\Lambda_{B_{Qqq}}$ 
and $\Lambda_{B_{qQQ}}$ and the constituent quark masses 
$m_u=m_d$, $m_s$, $m_c$ and $m_b$ have been taken from a fit in 
a previous analysis of the properties of single and double heavy 
baryons~\cite{Ivanov:1999bu,Faessler:2001mr}.   
In the present paper we have one remaining free parameter 
$\Lambda_{B_{QQQ}}$ characterizing the triple heavy baryons 
which will be fixed at 5 GeV using the heuristic 
relation for heavy baryons: 
$\Lambda_{B_{Qqq}} : \Lambda_{B_{qQQ}} : \Lambda_{B_{QQQ}} \simeq 
B_{Qqq} :  B_{qQQ} :  B_{QQQ}$, 
where $B_{Qqq}$, $B_{qQQ}$ and $B_{QQQ}$ are 
the typical masses of single, double and triple heavy 
baryons. The value of $\Lambda_{B_{QQQ}} = 5$ GeV is considered 
as preliminary because we need some data on triple heavy baryons. 
In Section 5 we also discuss the sensitivity of the magnetic moments 
of triple heavy baryons on the variation of the size parameter 
$\Lambda_{B_{QQQ}}$ in the region $3 - 7$ GeV.   

\section{Baryon Mass Operator and the baryon matrix element of the 
Electromagnetic current} 

We begin by expanding the baryon matrix element of the electromagnetic current 
in terms of the Dirac $F_{D}$ and 
the Pauli $F_{P}$ form factors:
\eq 
M_{\mu}&=&\bar{u}_B(p^{\prime}) \, \Lambda_{\mu}(p,p^{\prime}) 
u_B(p)\nonumber\\
\Lambda_{\mu}(p,p^{\prime})&=&\gamma_{\mu}F_{D}(q^2)+\frac{i}{2m_{B}}
\sigma_{\mu \nu}q^{\nu}F_{P}(q^2)\nonumber
\en 
where $u_B(p)$ is the baryon spinor with normalization  
$\bar u_B(p) \, u_B(p) = 2 m_B$. The momenta of the incoming photon, incoming 
and outcoming baryon are denoted, respectively, by 
$q$, $p$ and $p^\prime$ with $q = p^\prime - p$. 

The magnetic moment of the baryon is defined by 
\eq 
\mu_B \, = \, [ \, F_{D}(0) + F_{P}(0) \, ] \,\, \frac{e}{2 m_B} \,. 
\en 
We have set $\hbar = 1$. In terms of the nuclear magneton (n.m.) 
$\mu_N = \frac{e \, \hbar}{2 m_p}$ the baryon magnetic moment 
is given by 
\eq 
\mu_B  = 
\ [ \, F_{D}(0) + F_{P}(0) \, ] \,\, \frac{m_p}{m_B}  \quad\quad
{\rm (in \ units \ of \ n. m.)} 
\en 
where $m_p$ is the proton mass. 

We continue with a summary of some useful analytical results. 
The baryon mass operator is described by the Feynman diagram Fig.1. 
There are three diagrams that contribute to the electromagnetic 
vertex of the baryon: the triangle diagram Fig.3a 
and the two bubble diagrams Fig.3b and Fig.3c. In the triangle diagram 
the coupling of the photon to each of the three quark lines is implied. 
The bubble diagrams are generated by the nonlocal coupling of the photon
to the baryon made up of three quark fields as described by the 
the Lagrangian (\ref{gauging}) after expansion of the gauge exponential. 
Other possible diagrams at this order, where the photon couples
directly to the baryons, are excluded due to the compositeness 
conditions $Z_B = 0$ as discussed before.
 
In general for off-shell baryons, it is convenient to write down 
the electromagnetic vertex function for the transition 
$\Lambda_{\mu}(p,p^{\prime})$ in the form 
\eq\label{id1}
\Lambda_{\mu}(p,p^{\prime})=\frac{q_{\mu}}{q^{2}} \, e_B \, 
\biggl[ \Sigma_{B}(p^{\prime}) - \Sigma_{B}(p) \biggr] 
\, + \, \Lambda_{\mu }^{\perp}(p,p^{\prime}) 
\en 
where $\Lambda_{\mu }^{\perp}(p,p^{\prime})$ is 
the part of the vertex function which is orthogonal to the photon 
momentum $q^{\mu}\Lambda_{\mu}^{\perp}(p,p^{\prime})=0$. 
The explicit expression for $\Lambda^\mu_\perp(p,p^\prime)$ 
results from the sum of the gauge-invariant parts of the 
triangle $(\Delta)$ in Fig.3a and of the bubble $(\circ)$ diagrams 
in Figs.3b and 3c: 
\eq
\Lambda_{\mu}^{\perp}(p,p^{\prime}) = 
\Lambda_{\mu, \, \Delta}^{\perp}(p,p^{\prime}) + 
\Lambda_{\mu, \, \circ}^{\perp}(p,p^{\prime})
\en   
The separation (\ref{id1}) can be achieved in the following manner.
For the $\gamma$-matrices and four--vectors
we use the representation: 
\eq\label{split}
\gamma_\mu \, = \, \gamma_\mu^{\perp} \, + \, 
q_\mu \, \frac{\not\! q}{q^2}\,,  \hspace*{1cm}
k_\mu \, = \, k_\mu^{\perp} \, + \, q_\mu \, \frac{k q}{q^2}\,, 
\en 
such that $\gamma_\mu^\perp \,  q^\mu \, = \, 0$ and 
$k_\mu^\perp \, q^\mu=0$. The orthogonal vertex function 
$\Lambda_\mu^\perp(p,p^{\prime})$ is expressed in terms 
of $\gamma_\mu^\perp$ and $k_\mu^\perp$. 

Then the Ward-Takahashi identity~\cite{Ward:1950xp} relating 
the baryon electromagnetic vertex and the mass operator is 
satisfied according to  
\eq\label{id2} 
q^{\mu}\Lambda_{\mu}(p,p^{\prime}) = e_B \, 
\biggl[ \Sigma_{B}(p^{\prime}) - \Sigma_{B}(p) \biggr] \,.
\en 
In particular, if $q=0$ and both baryons are on their mass-shell, then 
the following Ward identity is satisfied: 
\eq\label{id3}
\Lambda_\mu(p,p) = e_B \, 
\frac{\partial}{\partial p_\mu}\Sigma_{B}(p) \,.
\en 
The identities (\ref{id2}) and (\ref{id3}) have to be satisfied in 
our approach since gauge invariance is fulfilled  
by construction of a gauge invariant Lagrangian.  

The expressions for the baryon mass operator $\Sigma_{B}(p)$ (Fig.1), 
the triangle $\Lambda_{\mu, \, \Delta}$ (Fig.3a) and the bubble diagrams 
$\Lambda_{\mu, \circ_L}^{\perp}$ (Fig.3b) and 
$\Lambda_{\mu, \circ_R}^{\perp}$ (Fig.3c) read 
(here and in the following we omit the flavor coefficients): 
\eq\label{elements}
\Sigma_{B}(p) &=& \alpha_B \, \int dk_{123} \,\, \tilde\Phi^{2}(z_0) 
\, R_{\Sigma}(k_1^+,k_2^+,k_3^+) \,, \nonumber\\[3mm]
\Lambda_{\mu, \, \Delta}^{\perp}(p,p^{\prime})
&=&\alpha_B \, \int dk_{123} \,\, \sum\limits_{i=1}^3 \, 
e_{q_i} \, \tilde\Phi(z_0) \, \tilde\Phi[z_0 + z_i(q)] 
\, R_{\mu \,, \Delta_i}^\perp(k_1^+,k_2^+,k_3^+,q)\,,  
\nonumber\\[3mm]
\Lambda_{\mu, \, \circ_L}^{\perp}(p,p^{\prime})
&=& - \alpha_B \, \int dk_{123} \,\, \sum\limits_{i=1}^3 \, 
e_{q_i} \, L_{i \mu}^\perp \, 
\tilde\Phi(z_0) \, \int\limits_0^1 \, dt \, 
\tilde\Phi^\prime[z_0 + t z_i(-q)]  
\, R_{\Sigma}(k_1^{\prime \, +},k_2^{\prime \, +},k_3^{\prime \, +}) 
\,,  \\[3mm]
\Lambda_{\mu, \, \circ_R}^{\perp}(p,p^{\prime})
&=& - \alpha_B \, \int dk_{123} \,\, \sum\limits_{i=1}^3 \, 
e_{q_i} \, L_{i \mu}^\perp \, 
\tilde\Phi(z_0) \, \int\limits_0^1 \, dt \, 
\tilde\Phi^\prime[z_0 + t z_i(q)] \, R_{\Sigma}(k_1^+,k_2^+,k_3^+) \, ,
\nonumber 
\en 
where for convenience we have introduced the notation: 
\eq
\alpha_B &=& 6 \, g_{B}^{2}\,, \quad k_i^+ \, = \, k_i + p \omega_i\,, 
\quad k_i^{\prime \, +} \, = \, k_i + p^\prime \omega_i\,, 
\quad z_0 \, = \,  - 6(k_{1}^{2}+k_{2}^{2}+k_{3}^{2})\, 
\nonumber\\[3mm]
dk_{123} &=& \frac{d^{4}k_{1}d^{4}k_{2}d^{4}k_{3}}{(2\pi)^8 i^2} 
\, \delta^{4}(k_{1}+k_{2}+k_{3})  \,, \quad 
L_i \, = \, 12 (k_i - \sum\limits_{j=1}^3 k_j \omega_j) \,, \\[3mm]
z_1(q) &=& - 12q^{2}(\omega_{2}^{2}+\omega_{2}\omega_{3}
+\omega_{3}^{2}) - L_1 q \,, \nonumber\\[5mm]
z_2(q) &=& - 12q^{2}(\omega_{1}^{2}+\omega_{1}\omega_{3}
+\omega_{3}^{2}) - L_2 q \,, \nonumber\\[5mm]
z_3(q) &=& - 12q^{2}(\omega_{1}^{2}+\omega_{1}\omega_{2} 
+\omega_{2}^{2}) - L_3 q \,. \nonumber  
\en 
and 
\eq 
R_{\mu \,, \, \Delta_1}^\perp(r_1,r_2,r_3,q) &=& 
\Gamma_{1f}S_{q_1}(r_1 + q)\gamma^{\perp}_{\mu}S_{q_1}(r_1)\Gamma_{1i} 
{\rm tr} \left[\Gamma_{2f}S_{q_2}(r_2)
\Gamma_{2i}S_{q_3}(-r_3)\right] \,, \nonumber\\[5mm] 
R_{\mu\,, \, \Delta_2}^\perp(r_1,r_2,r_3,q) &=& 
\Gamma_{1f}S_{q_1}(r_1)\Gamma_{1i} \, 
{\rm tr} \left[\Gamma_{2f}S_{q_2}(r_2 + q)
\gamma^{\perp}_{\mu} S_{q_2}(r_2)\Gamma_{2i} 
S_{q_3}(-r_3)\right] \,,\\[5mm]
R_{\mu\,, \, \Delta_3}^\perp(r_1,r_2,r_3,q) &=& 
- \Gamma_{1f}S_{q_1}(r_1)\Gamma_{1i} \, 
{\rm tr} \left[\Gamma_{2f}S_{q_2}(r_2)\Gamma_{2i} 
S_{q_3}(-r_3)\gamma^{\perp}_{\mu} S_{q_3}(-r_3 - q)\right] \,, 
\nonumber \\[5mm]
R_{\Sigma}(r_1,r_2,r_3) &=& \Gamma_{1f}S_{q_1}(r_1)\Gamma_{1i} \, 
{\rm tr} \left[\Gamma_{2f}S_{q_2}(r_2)\Gamma_{2i}S_{q_3}(-r_3)\right] \,. 
\nonumber 
\en 
The prime in Eq.~(\ref{elements}) on   
$\tilde\Phi^\prime$ denotes the derivative: 
\eq 
\tilde\Phi^\prime(s) \, = \, 
\frac{d\tilde\Phi(s)}{ds} \, .
\en 
Note that the expressions for the left (Fig.3a) and right (Fig.3b) 
bubble diagrams are related to each other via exchange of 
the external momenta $p \leftrightarrow p^\prime$ with  
\eq 
\Lambda_{\mu, \, \circ_L}^{\perp}(p,p^{\prime}) \, \equiv 
\Lambda_{\mu, \, \circ_R}^{\perp}(p^\prime,p) \,. 
\en 
In Appendix B we describe the calculation of the baryon mass operator 
and the vertex functions. 

\section{Heavy Quark Limit and Matching to Heavy Hadron ChPT} 

In this section we check the consistency of our approach with the 
model-independent predictions of Heavy Hadron Chiral Perturbation 
Theory (HHChPT) for the magnetic moments of single heavy baryons. 
Note, that HHChPT is the combination of Chiral Perturbation Theory 
(ChPT) and Heavy Quark Effective Theory (HQET) which is thus well suited 
for the description of the soft interactions of hadrons containing 
a single heavy quark with light pseudoscalar mesons and photons. 

In particular, HHCHPT predicts the following structure of the 
magnetic moments of the $\Lambda_Q$ and $\Sigma_Q$ baryons
including the leading and next-to-leading order terms in
$1/m_Q$ and $1/\Lambda_\chi$~\cite{Savage:1994zw,Banuls:1999mu}: 
\begin{eqnarray}\label{mu_HHCHPT} 
\mu_{\Lambda_{Q[q_2q_3]}} &=& \frac{e_Q}{2 m_Q} 
+ \frac{c_{\Lambda_{Qq_2q_3}}}{m_Q \, \Lambda_\chi} 
+ \frac{d_{\Lambda_{Qq_2q_3}}}{m_Q \, \Lambda_\chi^2} 
+ \ldots \,, \\ 
\mu_{\Sigma_{Q\{q_2q_3\}}} &=& - \frac{e_Q}{6 m_Q} 
+ \frac{c_{\Sigma_{Qq_2q_3}}}{\Lambda_\chi} 
+ \frac{d_{\Sigma_{Qq_2q_3}}}{\Lambda_\chi^2}   
+ \ldots \,, \nonumber 
\end{eqnarray}  
where $e_Q$ is the heavy quark charge, and $c_{B_{Qq_2q_3}}$ and $
d_{B_{Qq_2q_3}}$ 
are unknown coupling factors in the HHChPT approach. As we shall see further
on their numerical values can be determined in our approach. 
Here $\Lambda_{\chi} = 4 \pi F_\pi \simeq 1.2$ GeV is the scale
parameter of spontaneously broken chiral symmetry. It is known 
that the leading contribution ($e_Q/2m_Q$) to the magnetic moments 
of the $\Lambda_Q$ baryons comes from the coupling of the heavy quark 
to the photon. Therefore, $\mu_{\Lambda_{Q[q_2q_3]}}$ should vanish 
in the heavy quark limit. On the other hand, the leading contribution 
to the magnetic moment of the $\Sigma_Q$ type baryon, 
which survives in the heavy quark limit, comes 
from the coupling of the light quarks to the photon. The leading 
contribution to $\mu_{\Sigma_Q}$ due to the coupling of 
the heavy quark to the photon field is also proportional to $1/m_Q$ 
as in the case of the $\Lambda_Q$ type baryons. 
It should be clear that any phenomenological quark model should be able to 
reproduce these model independent predictions of HHChPT. 

Note that the terms proportional to the coupling factors  
$c_{\Lambda_{Qq_2q_3}}$, $d_{\Lambda_{Qq_2q_3}}$ and $d_{\Sigma_{Qq_2q_3}}$ 
originate from the meson-cloud (chiral) corrections. 
In the near future we intend to evaluate 
these corrections using the dressing formalism of 
the electromagnetic quark operator recently developed
in Ref.~\cite{Faessler:2005ah}. The dressing formalism is consistent with 
baryon ChPT~\cite{Gasser:1987rb}-\cite{Fuchs:2003qc}  
due to the matching of the physical amplitudes of both approaches 
at the baryonic level. In particular, we intend to 
perform a comprehensive analysis of the magnetic moments of 
light and heavy baryons including the contributions 
of valence and sea quarks. 

In the present manuscript we restrict ourselves to the contributions of the 
valence quark degrees of freedom to the magnetic moments of heavy baryons. 
We first reproduce the leading contributions to the 
$\Lambda_Q$ and $\Sigma_Q$ baryons and, second, estimate 
the coupling factors $c_{\Sigma_{Qq_2q_3}}$. 
 
To this end we expand the heavy quark propagator in powers 
of the inverse heavy quarks mass and keep only the leading term in 
the expansion. One has
\eq\label{S_Q} 
\tilde S_Q(k + p) = \frac{1}{m_Q - \not\! k - \not\! p} = 
- \frac{1 + \not\! v}{2 \, (kv + \bar\Lambda_{q_2q_3})} 
+ O(1/m_Q) 
\en 
where $v = p/m_B$ is the four-velocity of the single heavy baryon 
and $\bar\Lambda_{q_2q_3}$ is the difference between the masses of the 
single heavy baryon and the heavy quark in the heavy quark limit given by
\eq 
m_{B_{Qq_2q_3}} = m_Q + \bar\Lambda_{q_2q_3} + O(1/m_Q) \,. 
\en 
We use the same values of the 
$\bar\Lambda_{q_2q_3}$ parameters as in our previous 
papers~\cite{Ivanov:1996fj}-\cite{Faessler:2001mr}: 
\eq\label{Lambda_q2q3}
& &\bar\Lambda_{uu} = \bar\Lambda_{ud} = \bar\Lambda_{dd} = 
600 \ {\rm MeV}\,, \nonumber\\
& &\bar\Lambda_{us} = \bar\Lambda_{ds} = 750 \ {\rm MeV}\,, \\ 
& &\bar\Lambda_{ss} = 900 \ {\rm MeV}\,. \nonumber   
\en 
Using the calculational technique discussed in Appendix B 
we find that the contributions of the triangle diagram in Fig.3a 
(coupling of the photon to the heavy quark) to the magnetic moments 
of the $\Lambda$- and $\Sigma$-type baryons is in exact agreement 
with HHChPT:  
\eq\label{mu_Q} 
\mu_{\Lambda_{Q[q_2q_3]}}^{\rm heavy} &=&   \frac{e_Q}{2 m_Q} \, \\
\mu_{\Sigma_{Q\{q_2q_3\}}}^{\rm heavy}  &=& - \frac{e_Q}{6 m_Q} \,. 
\en  
These results are independent of the form of the 
baryon correlation function and of the flavor content 
of the light diquark system. 

We stress that the bubble diagrams in Figs.3b and 3c  
do not contribute to the magnetic moment at the order of accuracy that we 
are interested in. 
For the contribution of the triangle diagrams describing the coupling of the 
photon to the light quarks we obtain the following results. 
In agreement with HHChPT there is no contribution to 
to magnetic moments of $\Lambda$-type baryons at order $O(1/m_Q)$. 
The leading contribution to $\mu_{\Sigma_{Q\{q_2q_3\}}}$ 
which survives in the heavy quark limit is given by 
\eq 
\mu_{\Sigma_{Q\{q_2q_3\}}}^{\rm light} \, = \, 
\frac{1}{\Lambda_{B_Q}} \, 
\frac{I_2(m_{q_2},m_{q_3})}{I_1(m_{q_2},m_{q_3})}
\en 
where the integrals $I_1$ and $I_2$ depend on the model parameters 
(the baryon correlation function and the light flavors through their 
constituent quark masses $m_{q_2}$ and $m_{q_3}$): 
\eq\label{I12}  
I_1(m_{q_2},m_{q_3}) &=& 3 \, (e_Q + e_{q_2} + e_{q_3}) \,  
\int\limits_0^\infty \int\limits_0^\infty \int\limits_0^\infty \, 
d\alpha_1 \, \frac{d\alpha_2 d\alpha_3}{[{\rm det}A]^2} \, 
\tilde\Phi^2(12z) \, \biggl[ \mu_{q_2} \mu_{q_3} 
+ \frac{\alpha_1^2}{{\rm det}A} 
- \frac{\bar\lambda \alpha_1}{2{\rm det}A} \biggr] \,, \\
I_2(m_{q_2},m_{q_3}) &=&  
\int\limits_0^\infty \int\limits_0^\infty \int\limits_0^\infty \, 
d\alpha_1 \alpha_1 \, \frac{d\alpha_2 d\alpha_3}{[{\rm det}A]^2} \, 
\tilde\Phi^2(12z) \, (\alpha_2 e_{q_2} + \alpha_3 e_{q_3}) \, 
\biggl[ \mu_{q_2} \mu_{q_3} 
+ \frac{\alpha_1^2}{{\rm det}A} 
\biggl( 1  + \frac{1 + \alpha_{23}}{2{\rm det}A}  \biggr) 
- \frac{\bar\lambda \alpha_1}{{\rm det}A} \biggr] \nonumber 
\en 
where 
\eq 
& &\mu_{q_i} = \frac{m_{q_i}}{\Lambda_{B_Q}}\,, \quad 
\bar\lambda = \frac{\bar\Lambda}{\Lambda_{B_Q}}\,, 
\quad \alpha_{ij} = \alpha_i + \alpha_j \,, 
\quad {\rm det}A = \frac{3}{4} + \alpha_{23} + 
\alpha_2 \alpha_3 \, \\
& &z = \mu_{q_2}^2 \alpha_2 + \mu_{q_2}^3 \alpha_3 
+ \frac{\alpha_1^2}{{\rm det}A} (1 + \alpha_{23})  
- 2 \bar\lambda \alpha_1 \, 
\nonumber 
\en 
and 
$\Lambda_{B_Q}$ is the size parameter appearing in the correlation function 
of the single heavy baryons $\tilde\Phi(z)$. 

Our prediction for the unknown HHChPT coupling 
$c_{\Sigma_{Qq_2q_3}}$ is 
\eq\label{c_Sigma} 
c_{\Sigma_{Qq_2q_3}} \, = \, \frac{\Lambda_\chi}{\Lambda_{B_Q}} \,  
\frac{I_2(m_{q_2},m_{q_3})}{I_1(m_{q_2},m_{q_3})} \,. 
\en 
We complete our analysis of the magnetic moments of the  
$\Lambda_Q$-- and $\Sigma_Q$--type baryons by deriving the 
magnetic moments $\mu_{\Lambda_Q}$ and $\mu_{\Sigma_Q}$ 
in the framework of the naive nonrelativistic quark model. We use 
the baryonic spin-flavor wave functions 
(\ref{LambdaQ_wf}) and (\ref{SigmaQ_wf}) arising from the 
relativistic three-quark currents in the 
nonrelativistic limit. After some simple algebra we get 
\eq\label{mu_NRQM}
\mu_{\Lambda_{Q[q_2q_3]}}  & = &  
\la \Lambda_{Q[q_2q_3]} | \sum\limits_{i=1}^3 \frac{e_{q_i}}{2m_{q_i}} 
\, \sigma_{3i} | \Lambda_{q_1[q_2q_3]} \ra \, = \, 
\frac{e_Q}{2m_Q} \,, \\
\mu_{\Sigma_{Q\{q_2q_3\}}} & = &  
\la \Sigma_{Q\{q_2q_3\}} | 
\sum\limits_{i=1}^3 \frac{e_{q_i}}{2m_{q_i}} 
\, \sigma_{3i} | \Sigma_{Q\{q_2q_3\}} \ra \, = \, 
- \frac{e_Q}{6m_Q} + \frac{e_{q_2}}{3m_{q_2}} + \frac{e_{q_3}}{3m_{q_3}} 
\,. \nonumber 
\en 

\section{Numerical Results and Discussion}

Our results for magnetic moments of heavy baryons are given 
in Tables 4-6. 
First, we discuss the numerical results for the magnetic moments 
of single heavy baryons for three models (see Table 4): 
1) exact results (full) without any approximations; 2) heavy 
quark limit (HQL) for the baryonic correlation function (BCF)  
and 3) full HQL for both the BCF and the heavy quark propagator (HQP).  
We refer to these models as full, HQL BCF and HQL BCF+HQP, 
respectively. For the convenience we separate the 
contributions coming from the coupling of the heavy quark ($c$ or $b$) 
to the photon (heavy quark contribution) and the same for the light 
degrees of freedom (light quark contribution). In the numerical 
calculations we use the Gaussian form of the baryonic correlation 
function~(\ref{Gauss_CF}). The last column contains the results 
of the nonrelativistic quark model (NRQM) which uses the spin-flavor 
baryonic wave functions described in the previous sections. 
The basic notions of the NRQM are given in Appendix C. 
In particular, in Table 7 we present the wave functions 
and magnetic moments of heavy baryons in the NRQM. 

Next we predict the unknown HHChPT couplings $c_{\Sigma_{Qq_2q_3}}$ 
using Eqs.~(\ref{mu_HHCHPT}) and (\ref{c_Sigma}) restricting ourselves to 
the model HQL BCF+HQP for $\Sigma$-type baryons. 
For the sake of comparison we following Ref.~\cite{Banuls:1999mu} in the 
construction of the flavor coefficients in $c_{\Sigma_{Qq_2q_3}}$. One has 
\eq 
c_{\Sigma_{Qq_2q_3}} \ = \ \frac{4}{9} \ c_S \ \mu_{\Sigma_{Qq_2q_3}}^S 
\en    
where 
\eq 
& &\mu_{\Sigma_c^{++}}^S = \mu_{\Sigma_b^{+}}^S = 2\,, \quad 
\mu_{\Sigma_c^{+}}^S = \mu_{\Sigma_b^{0}}^S = \frac{1}{2}\,, \quad  
\mu_{\Sigma_c^{0}}^S = \mu_{\Sigma_b^{-}}^S = - 1\,, \\ 
& &\mu_{\Xi_c^{^\prime 0}}^S = \mu_{\Xi_b^{^\prime -}}^S = - 1\,, \quad  
\mu_{\Xi_c^{^\prime +}}^S = \mu_{\Xi_b^{^\prime 0}}^S = \frac{1}{2} 
\,, \quad  \mu_{\Omega_c^{0}}^S = \mu_{\Omega_b^{-}}^S = -1 \; . \nonumber 
\en  
In Table 5 we present our predictions for the 
HHChPT coupling factor $c_S$. In our approach the value of this coupling 
factor depends on the light quark content of the heavy baryon. It is also 
flavor dependent because we break SU(3) flavor symmetry. 
Therefore, there are different predictions (see Table 5) for 
the $c_S$ coupling factors depending on whether one is dealing with 
nonstrange, single strange or double strange states. 
Even for the cascade states $\Xi_{Q\{us\}}^\prime$ and 
$\Xi_{Q\{ds\}}^\prime$ we get different predictions, because the 
contributions of $u$ and $d$-quarks enter with different coefficients. 
While the coupling factors $c_S$ do not depend on the heavy flavor 
(as stressed in Ref.~\cite{Banuls:1999mu}) they do depend on the light 
quark flavor. The coupling factors $c_S$ in Table~5 vary from 0.26 to 0.55
depending on the $SU(3)$ flavor content. 
In Table 6 we present our results for the magnetic moments of double and 
triple heavy baryons for two models: 1) exact calculation (full); 
2) heavy quark 
limit (HQL) for the baryonic correlation function (BCF). 
We also perform a comparison with the results of the NRQM. 

Finally we compare our predictions for the magnetic moments of 
heavy baryons with the results of other theoretical approaches: 
QCD sum rules~\cite{Zhu:1997as,Aliev:2001ig}, 
soliton approaches~\cite{Oh:1991ws}-\cite{Scholl:2003ip} 
and quark models~\cite{Bose:1980vy,Glozman:1995xy,Julia-Diaz:2004vh}.  
First of all, we stress again that in the sector of single heavy 
baryons we are consistent with HHChPT~\cite{Banuls:1999mu} in the 
heavy quark limit. We reproduce the leading terms in the expansion 
of the magnetic moments in powers of $1/m_Q$ and $1/\Lambda_\chi$. 
Moreover, we were able to predict the unknown 
HHChPT coupling factors $c_S = 0.26 - 0.55$. 

The magnetic moments of $\Lambda$-type single heavy baryons are practically 
the same in all three models, because the leading contribution 
comes from the coupling of the photon to the heavy quark. This is not the case 
for the $\Sigma$-type baryons: the contribution of the heavy quarks 
remains unchanged but the light quark contribution is increased 
in the ``full'' scheme while it is suppressed in the HQL BCF and HQL BCF+HQP 
schemes.  
We mention that the contribution of the bubble diagrams Fig.3b and Fig.3c is 
suppressed. In magnitude it is less than 5\%.    
Note, that the predictions of the HQL BCF and HQL BCF+HQP models are 
very similar.  

Our ``full'' model is close to the predictions of the naive quark 
model as well as the results obtained in the quark 
models~\cite{Jena:1986xs,Glozman:1995xy} given by~\cite{Jena:1986xs}  
\eq 
& &\mu_{\Lambda_c^+} = \mu_{\Xi_c^+} = \mu_{\Xi_c^0} = 0.35\, 
\nonumber\\ 
& &\mu_{\Sigma_c^{++}} = 2.37-2.45\,, \quad 
   \mu_{\Sigma_c^+}    = 0.50-0.52\,,   \quad 
   \mu_{\Sigma_c^0}    = - (1.36-1.40)\, \nonumber\\  
& &\mu_{\Xi_c^{\prime +}} = 0.75-0.78\,, \quad 
   \mu_{\Xi_c^{\prime 0}}  = - (1.12-1.15) \,, \quad 
   \mu_{\Omega_c^0}   = - (0.88-0.89)\, \\
& &\mu_{\Sigma_b^+} = 2.50-2.59\,, \quad 
   \mu_{\Sigma_b^0} = 0.64-0.66\,,   \quad 
   \mu_{\Sigma_b^-} = - (1.22-1.26)\, \nonumber\\  
& &\mu_{\Xi_b^{\prime 0}} = 0.88-0.92\,, \quad 
   \mu_{\Xi_b^{\prime -}}  = - (0.98-1.01) \,, \quad 
   \mu_{\Omega_b^-}  = - (0.74-0.75)\, \nonumber
\en 
and~\cite{Glozman:1995xy}
\eq 
& &\mu_{\Lambda_c^+} = \mu_{\Xi_c^+} = \mu_{\Xi_c^0} = 0.38\, 
\nonumber\\ 
& &\mu_{\Sigma_c^{++}} = 2.33\,, \quad \mu_{\Sigma_c^+}  = 0.49\,, 
\quad \mu_{\Sigma_c^0}  = - 1.35\, \\  
& &\mu_{\Xi_c^{\prime +}} = 0.65\,, \quad 
\mu_{\Xi_c^{\prime 0}}  = - 1.18 \,, 
\quad \mu_{\Omega_c^0}  = - 1.02\, \nonumber
\en 
The ``full'' scheme results can be compared to the 
predictions of the QCD sum rule approach~\cite{Zhu:1997as}: 
\eq 
\mu_{\Sigma_c^{++}} = 2.1  \pm 0.3\,,  \quad 
\mu_{\Sigma_c^+}    = 0.6  \pm 0.1\,,  \\ 
\mu_{\Sigma_c^{0}}  = - (1.6 \pm 0.2)\,,  \quad 
\mu_{\Lambda_c^+}    = 0.15 \pm 0.05\, \nonumber 
\en 
and~\cite{Aliev:2001ig}: 
\eq 
\mu_{\Lambda_c^+}    = 0.40 \pm 0.05\,, \quad\quad 
\mu_{\Lambda_b^0}    = - (0.18 \pm 0.05)\,. 
\en 
Let us stress that the agreement 
with the NRQM is based on the use of the specific spin-flavor wave functions 
of baryons which correspond to our relativistic baryonic currents 
in the nonrelativistic limit. The use of other NRQM spin-flavor 
structures give very different results.  

A detailed analysis of the magnetic moments of single heavy baryons 
has been performed in solitonic (Skyrme) 
approaches~\cite{Oh:1991ws,Oh:1995eu,Scholl:2003ip}. 
The most recent calculation~\cite{Scholl:2003ip} gives the following numbers 
for $\Lambda$-type baryons  
\eq 
\mu_{\Lambda_c^+}   = 0.12 -0.13 \,, \quad\quad 
\mu_{\Lambda_b^0}    = -0.02\,,  
\en  
which are smaller than our numbers. 
Their results for the $\Sigma$-type baryons are  
\eq 
\mu_{\Sigma_c^{++}} = 2.45 - 2.46 \,,  \quad 
\mu_{\Sigma_c^0}    = -1.96\,,  \\ 
\mu_{\Sigma_b^+}  = 2.52\,,  \quad 
\mu_{\Sigma_b^-}    = - (1.93 - 1.94)\,. \nonumber 
\en 
are larger than ours in average.
Recently, the magnetic moments of charmed baryons have 
been calculated in a relativistic quark model~\cite{Julia-Diaz:2004vh} 
exploiting three different forms of relativistic kinematics. 
In the case of single heavy baryons it was found that there is only a small 
dependence on the kinematics for $\Lambda$-type baryons. For the
magnetic moments of the $\Lambda$-type baryons they quote:  
\eq 
\mu_{\Lambda_c^+} = 0.39 - 0.52\,, \quad 
\mu_{\Xi_c^+} = 0.39 - 0.47\,, \quad  
\mu_{\Xi_c^0} = 0.39 - 0.47\,. 
\en   
Contrary to this there is a strong dependence on 
the form of the relativistic kinematics for $\Sigma$-type baryons. For these
they quote  
\eq 
\mu_{\Sigma_c^{++}} = 0.90 - 3.07\,, \quad  
\mu_{\Sigma_c^{0}} = - (0.74 - 1.78)\,, \quad 
\mu_{\Omega_c^{0}} = - (0.67 - 1.03)\,.  
\en 
Our predictions in both our HQL schemes are similar 
to the results of the MIT bag model~\cite{Bose:1980vy}, 
for which one obtains  
\eq 
\mu_{\Sigma_c^{++}} = 0.70\,, \quad 
\mu_{\Sigma_c^{0}} = - 0.44\,, \quad 
\mu_{\Omega_c^{0}} = - 0.35\,, \\ 
\mu_{\Sigma_b^{+}} = 0.83\,, \quad 
\mu_{\Sigma_b^{-}} = - 0.40\,, \quad 
\mu_{\Omega_b^{-}} = - 0.30\,. \nonumber 
\en  
Now we turn to the results for the magnetic moments 
of double and triple heavy baryons (see Table 6). 
One can see that the results of the ``full'' model are close to the 
results of the NRQM. Note that the predictions for the HQL BCF 
model are similar to the results of the relativistic 
quark model~\cite{Julia-Diaz:2004vh} if one uses the ``point'' 
form of the relativistic kinematics, e.g.: 
\eq 
\mu_{\Xi_{cc}^{++}} = 0.29 - 0.30\,, \quad 
\mu_{\Xi_{cc}^{+}}  = 0.68 - 0.69\,, \quad 
\mu_{\Omega_{cc}^{+}} = 0.66\,. 
\en   
Also there is an agreement for some states with the relativistic 
quark potential model~\cite{Jena:1986xs}: 
\eq 
& &\mu_{\Xi_{cc}^+} = 0.78-0.79\,, \quad 
\mu_{\Omega_{cc}^+}  = 0.66\,, \quad 
\mu_{\Xi_{bb}^0} = -(0.71-0.73)\,, \\
& &\mu_{\Xi_{bb}^-} = 0.23-0.24\,, \quad 
\mu_{\Omega_{bb}^-}  = 0.11\,, \quad 
\mu_{\Xi_{cb}^+} = 1.50-1.54\,. 
\en  
We found the agreement with Ref.~\cite{Jena:1986xs} 
for the magnetic moments of triple heavy baryons: 
\eq 
\mu_{\Omega_{ccb}^+} = 0.49\,, \quad\quad 
\mu_{\Omega_{cbb}^0} = -0.20 \,.
\en  
Finally, we show the sensitivity of the magnetic moments of triple 
heavy baryons to a variation of the size parameter 
$\Lambda_{B_{QQQ}}$ in the region $3 - 7$ GeV. When the value of 
$\Lambda_{B_{QQQ}}$  is varied from $3$ to $7$ GeV the values of 
$\mu_{\Omega_{ccb}^+}$ and $\mu_{\Omega_{cbb}^0}$ are changed as: 
\eq 
\mu_{\Omega_{ccb}^+} = 0.58 - 0.50 \,, \quad\quad 
\mu_{\Omega_{cbb}^0} = - (0.21 - 0.20)\,  
\en 
in the ``full'' model and 
\eq 
\mu_{\Omega_{ccb}^+} = 0.09 - 0.16 \,, \quad\quad 
\mu_{\Omega_{cbb}^0} = - (0.11 - 0.14 )\, 
\en 
in the HQL BCF scheme.   

It will be interesting to compare our result for the magnetic moments 
of double and triple heavy baryons with possible future results in the  
framework of nonrelativistic QCD (NRQCD) recently extended on 
the sector of baryons containing two and three heavy 
quarks~\cite{Brambilla:2005yk,Fleming:2005pd}. 

\section{Conclusion}

We have employed the relativistic three-quark model 
to calculate the magnetic moments of single, double 
and triple heavy baryons. We have used a Gaussian shape for the 
baryon-quark vertex. For the propagators we have used free quark 
propagators. The electromagnetic 
vertex functions of the heavy baryons are described by a set of 
three-quark (triangle and bubble) diagrams. 
The parameters of the model are the constituent 
quark masses and the size parameters $\Lambda_{B}$ which appear as
free parameters in the baryonic 
correlation functions. 
We have presented a detailed analysis of the magnetic moments of heavy 
baryons. We have shown that our results have the correct structure
predicted by the model independent approach of HHChPT. This allowed us
to fix the values of some of the coupling factors that appear in HHChPT. 
Finally, we have compared our numerical results to the results of a variety
of other approaches. 

\begin{acknowledgments}
This work was supported by the by the DFG under contracts FA67/25-3 and
GRK683. This research is also part of the EU Integrated Infrastructure
Initiative Hadron physics project under contract number RII3-CT-2004-506078
and President grant of Russia ``Scientific Schools'' No. 5103.2006.2. 
M.A.I. also appreciates the partial support by 
the Russian Fund of Basic Research under Grant No. 04-02-17370.  
K.P. thanks the Development and Promotion of Science and Technology 
Talent Project (DPST), Thailand for financial support.  
\end{acknowledgments} 

\newpage 

\appendix 
\section{Feynman rule for the nonlocal electromagnetic vertex}  

In the following we derive the Feynman rules for the nonlocal vertex 
of Fig.1 describing the coupling of a baryon, the three quarks and the photon 
field. 
This vertex contains the path integral over the gauge field 
\eq 
I(x,y,P) = \int\limits_y^x d z_\mu A^\mu(z) \,. 
\en 
The crucial point is to calculate the expression 
\eq\label{MB3qg0}
\Gamma_{B3q\gamma} = \int d^4 x_1 \int d^4 x_2 \, 
\Phi(x_1^2 + x_2^2) \,  e^{ip_1x_1 + ip_2x_2} I(x_+,x,P) 
\en  
where $x_+ = x \pm a_1x_1 \pm a_2x_2$ and where $p_1$ and $p_2$ are 
linear combination of the momenta. 
We do not specify the parameters 
$a_{1,2}$ and momenta $p_{1,2}$ because their specific form is not 
necessary for the further derivation.     

We use the operator identity  
\eq\label{Oper_Id} 
\Phi(x_1^2 + x_2^2) 
&=& \int \frac{d^4 k_1}{(2\pi)^4} \, \int \frac{d^4 k_2}{(2\pi)^4} \, 
\tilde\Phi(-k_1^2 -k_2^2) \, e^{i k_1 x_1 + i k_2 x_2} \,, 
\nonumber\\ 
&=&\int \frac{d^4 k_1}{(2\pi)^4} \, \frac{d^4 k_2}{(2\pi)^4} \, 
\tilde\Phi(\partial_{x_1}^2 + \partial_{x_2}^2) \, 
e^{i k_1 x_1 + i k_2 x_2}  = \delta^4(x_1) \, \delta^4(x_2) \, 
\tilde\Phi(\partial_{x_1}^2 + \partial_{x_2}^2)   \,,     
\en 
where $\tilde\Phi$ is the Fourier transform of the vertex function 
$\Phi$. We then get  
\eq\label{MB3qg}
\Gamma_{B3q\gamma} = \int d^4 x_1 \int d^4 x_2 \,\,  
\delta^4(x_1) \,\, \delta^4(x_2) \,\,  
\tilde\Phi(\partial_{x_1}^2 + \partial_{x_2}^2) \,  
e^{ip_1x_1 + ip_2x_2} I(x_+,x,P) \,. 
\en 
It is readily seen that 
\eq 
\tilde\Phi(\partial_{x_1}^2 + \partial_{x_2}^2) 
e^{ip_1x_1 + ip_2x_2} I(x_+,x,P) = e^{ip_1x_1 + ip_2x_2} 
\tilde\Phi({\mathcal D}_{x_1}^2+{\mathcal D}_{x_2}^2) I(x_+,x,P) 
\en 
where ${\mathcal D}_{x_i} \equiv \partial_{x_i} + i p$.    

In order to evaluate
\eq 
\tilde\Phi({\mathcal D}_{x_1}^2+{\mathcal D}_{x_2}^2) I(x_+,x,P) 
\, = \, \sum\limits_{n=0}^{\infty} \, 
\frac{\tilde\Phi^{(n)}(0)}{n!} \, 
[{\mathcal D}_{x_1}^2+{\mathcal D}_{x_2}^2]^n \, I(x_+,x,P) \,  
\en 
we make use of the key equation~(\ref{path2}). One obtains 
\eq 
\partial_x^\mu I(x,y,P) = A^\mu(x) \, ,
\en 
resulting in
\eq 
[{\mathcal D}_{x_1}^2+{\mathcal D}_{x_2}^2] I(x_+,x,P) \, = \, 
L(A) - (p_1^2 + p_2^2) I(x_+,x,P) \; , 
\en 
where 
\eq 
L(A) \equiv (\partial_{x_1} + \partial_{x_2}) A(x) 
+ 2i (p_1 + p_2) A(x) \,. 
\en 
The iteration of the last expression gives the sequence
\eq 
( {\cal D}_{x_1}^2 + {\mathcal D}_{x_2}^2 )^2 I(x_+,x,P) 
&=& [{\mathcal D}_{x_1}^2+{\mathcal D}_{x_2}^2 
- (p_1^2 + p_2^2)] L(A) 
+ (-p_1^2 - p_2^2)^2 I(x_+,x,P) \,, \\
( {\mathcal D}_{x_1}^2 + {\mathcal D}_{x_2}^2 )^3 I(x_+,x,P) 
&=& [ ({\mathcal D}_{x_1}^2+{\mathcal D}_{x_2}^2)^2 - 
({\mathcal D}_{x_1}^2 + {\mathcal D}_{x_2}^2) (p_1^2 + p_2^2) 
+ (- p_1^2 - p_2^2)^2] L(A) 
+ (- p_1^2 - p_2^2)^3 I(x_+,x,P)  \,, \nonumber\\ 
&\ldots&\nonumber\\
( {\mathcal D}_{x_1}^2 + {\mathcal D}_{x_2}^2 )^n I(x_+,x,P) 
&=& \sum\limits_{k=0}^{n-1} \, 
({\mathcal D}_{x_1}^2+{\mathcal D}_{x_2}^2)^{n-1-k} \, 
(- p_1^2 - p_2^2)^k \, L(A) \, + \, ( - p_1^2 - p_2^2)^n 
I(x_+,x,P) \nonumber\\
&=& n \, \int\limits_0^1 \, dt \,\,  
[({\mathcal D}_{x_1}^2+{\mathcal D}_{x_2}^2) t \, - \, 
(p_1^2 +p_2^2) (1-t) ]^{n-1} \, L(A) \, + \, 
( - p_1^2 - p_2^2)^n \, I(x_+,x,P)\, . 
\nonumber 
\en 
One finally has
\eq\label{EqN1_last}
\tilde\Phi({\mathcal D}_{x_1}^2+{\mathcal D}_{x_2}^2) I(x_+,x,P) 
&=& \int\limits_0^1 \, dt \, 
\tilde\Phi^\prime [ ({\mathcal D}_{x_1}^2+{\mathcal D}_{x_2}^2)t 
\, -  \, (p_1^2 + p_2^2) (1-t) ] \, L(A) \, + 
\, \tilde\Phi(-p_1^2 - p_2^2) \, I(x_+,x,P) \nonumber\\ 
&=&\int \frac{d^4 q}{(2\pi)^4} \,  \tilde A_\mu(q) \, 
\bigg\{ i K^\mu  e^{- i q x_+ } \int\limits_0^1 dt 
\tilde\Phi^\prime[w(t)] + \tilde\Phi [w(0)] 
\int\limits_x^{x_+} dz^\mu e^{- iqz} \biggr\} \,,   
\en 
where 
$$K^\mu = a_1 [2 p_1 - q]^\mu + a_2 [2 p_2 - q]^\mu \,,$$ 
$$w(t) = - (p_1 - a_1 q)^2 t 
- (p_2 - a_2 q)^2 t - (p_1^2 + p_2^2) (1-t)\,,$$ 
$\tilde A_\mu(q)$ is the Fourier-transform of 
the electromagnetic field and 
$\tilde\Phi^\prime(z) = d\tilde\Phi(z)/dz$. 
The last term in Eq.(\ref{EqN1_last}) contains an integration 
from $x$ to $x_+$ in the path integral. This term vanishes due to the 
delta functions $\delta^4(x_1)$ and $\delta^4(x_2)$ in Eq.~(\ref{MB3qg}).  

Finally, we get
\eq\label{MB3qgf}
\Gamma_{B3q\gamma} = i \, \int \frac{d^4 q}{(2\pi)^4} \, e^{- i q x} \, 
 \tilde A_\mu(q) \, K^\mu \, \int\limits_0^1 dt \, 
\tilde\Phi^\prime[w(t)] \,.  
\en

\section{Details of the calculation of matrix elements} 

As an example of the evaluation of matrix elements we explicitly
calculate the baryon mass operator of Eq.(\ref{elements}). 
The generic integral reads: 
\eq\label{Pi_tech}
I_B(p) \, = \, \int dk_{123} \,\, \tilde\Phi^{2}(z_0) 
\, R_{\Sigma}(k_1^+,k_2^+,k_3^+) \,. 
\en 
For simplicity we set $w_1 = 1$ and 
$w_2 = w_3 = 0$ in Eq.~(\ref{Pi_tech}). 
Then the integral $I_B(p)$ can be written as: 
\eq 
I_B(p) = \int \frac{d^4k_1}{\pi^2 i}  \, 
\int \frac{d^4k_2}{\pi^2 i} \,\, \tilde\Phi^{2}(-12[k_1^2+k_1k_2+k_2^2]) 
\, \Gamma_{1f} \, S_{q_1}(k_1 + p) \, \Gamma_{1i} 
{\rm tr} \left[\Gamma_{2f} \, S_{q_2}(k_2) \, \Gamma_{2i} \, 
S_{q_3}(k_1+k_2)\right] \,. 
\en 
The technique we use is based on the following main ingredients:  
\begin{itemize}
\item  use of the Laplace transform of the vertex function, its 
derivative and integral: 
$$
\tilde\Phi(z_0)=\int\limits_0^\infty\! ds\, \Phi_L(s)\, e^{-sz_0} \,,
$$ 
$$
\tilde\Phi^\prime(z_0) = - \int\limits_0^\infty\! ds\, s \, 
\Phi_L(s)\, e^{-sz_0} \,,
$$ 
$$
\int\limits_0^\infty d\alpha \, \alpha^n \, 
\tilde\Phi(z_0+\alpha) = \Gamma(n+1) \, 
\int\limits_0^\infty\! \frac{ds}{s^n} \, 
\Phi_L(s)\, e^{-sz_0} \,,
$$ 

\vspace{-0.2cm}
\item  $\alpha$-transform of the propagator functions
$S_{q_1}$, $S_{q_2}$ and $S_{q_3}$
$$
\frac{1}{m_q^2-(k+p)^2}=\int\limits_0^\infty\! d\alpha \,  
e^{-\alpha (m^2-(k+p)^2)} \, ,
$$
\vspace{-0.2cm}
\item  differential representation of the numerator
$$
\left(\,m+\not\! k +\not\! p \right)\, e^{kq}=
\left(\,m+\gamma^\mu \frac{\partial}{\partial q^\mu} 
+ \not\! p\,\right)\, e^{kq} \, .
$$
\vspace{-0.2cm}
\item  Gaussian integral over virtual momenta $k_1$ and 
$k_2$
$$
\prod\limits_{j=1}^n \int \frac{d^4k_j}{\pi^2 i} 
\, \exp[ \, kAk + 2Bk \, ] = \frac{1}{{[\rm det} A]^n} 
\exp[ \, - B \, A^{-1} \, B \, ]
$$ 
where, in a general approach, $A$ is a $n \times n$ matrix and $B$ a 
n-component vector. In the present application we have $n=2$.  
\end{itemize} 

After some algebra we get the following expressions for 
the structure integral $I_B(p)$ 
\eq 
I_B(p) &=& \int\limits_0^\infty d\alpha_1 
\int\limits_0^\infty d\alpha_2  \int\limits_0^\infty d\alpha_3 
\int\limits_0^\infty d\beta \,\, \tilde\Phi^2[-12(z+\beta)]  \, 
\nonumber\\ 
&\times&\biggl\{ \, \Gamma_{1f} \, D_1 \, \Gamma_{1i} 
{\rm tr} \left[\Gamma_{2f} \, D_2 \, \Gamma_{2i} \, D_3 \right] 
- \beta \, \Gamma_{1f} \, D_1 \, \Gamma_{1i} 
{\rm tr} \left[\Gamma_{2f} \, \gamma^\mu \, 
\Gamma_{2i} \, \gamma_\mu  \right] (A^{-1}_{12} + A^{-1}_{22}) \\
&-& \beta \, \Gamma_{1f} \, \gamma^\mu \, \Gamma_{1i} 
{\rm tr} \left[\Gamma_{2f} \, \gamma_\mu \, \Gamma_{2i} \, 
D_3  A^{-1}_{12} + \Gamma_{2f} \, D_2 \, \Gamma_{2i} \, 
\gamma_\mu (A^{-1}_{11} + A^{-1}_{12}) \right] \biggr\}  \,. 
\en 
where 
\eq  
& & D_i \, = \, m_{q_i} + \not\! P_i\,, \quad 
    P_1 \, = \, p - B_1 A^{-1}_{11}\,, \quad 
    P_2 \, = \, - B_1 A^{-1}_{12}\,, \quad \\
& & P_3 \, = \,  - B_1 (A^{-1}_{11} + A^{-1}_{12}) \,, \quad 
    z \, = \, - \sum\limits_{i=1}^3 \alpha_i m_{q_i}^2 
    + p^2 \alpha_1 - B_1^2 A^{-1}_{11} \,. \nonumber 
\en 
Here $B_1 = p \alpha_1$ and 
$A^{-1}_{ij}$ are the elements of the inverse matrix $A_{ij}$: 
\eq 
A = 
\left(
\begin{array}{llll} 
1 + \alpha_1 + \alpha_3 & & & \frac{1}{2} + \alpha_3  \\
\frac{1}{2} + \alpha_3  & & & 1 + \alpha_2 + \alpha_3 \\
\end{array}
\right); \hspace*{1cm} 
A^{-1} = \frac{1}{{\rm det}A}  
\left(
\begin{array}{llll} 
1 + \alpha_2 + \alpha_3     & & & -(\frac{1}{2} + \alpha_3)  \\
- (\frac{1}{2} + \alpha_3)  & & & 1 + \alpha_1 + \alpha_3 \\
\end{array}
\right). 
\en 
The advantage of choosing this order of integration (made possible
by the use of Laplace transforms) is that the
specification of the baryonic correlation function 
$\tilde\Phi$ can be left to the last step after having integrated
over the virtual momenta.
All further calculations have been done by using computer programs written 
in FORM for the manipulations of the Dirac matrices and in FORTRAN
for the final numerical evaluation.

\section{Nonrelativistic quark model: spin-flavor wave functions and 
magnetic moments}

In this Appendix we present the results for the magnetic moments 
of heavy baryons within the nonrelativistic quark model. As emphasized before 
the nonrelativistic quark model is based on the spin-flavor wave functions 
which arise in the nonrelativistic limit of the relativistic covariant 
three-quark currents carrying the quantum numbers of $\frac{1}{2}^+$ baryons. 
Using Eqs.~(\ref{LambdaQ_wf}) and (\ref{SigmaQ_wf}) we specify 
the wave functions of all baryonic states involved in our calculations. 
Then we derive the expressions for the baryonic magnetic moments using 
the master formula: 
\eq 
\mu_{B_{q_1q_2q_3}} \, = \,   
\la B_{q_1q_2q_3} | \sum\limits_{i=1}^3 \frac{e_{q_i}}{2m_{q_i}} 
\, \sigma_{3i} | B_{q_1q_2q_3} \ra \,  
\en   
where $\sigma_{3i}$ is the third component of the spin operator of the 
$i$-th quark. 

In Table 7 we display our results for the wave functions and our 
predictions for the magnetic moments, where we use the following
notation for the antisymmetric 
$\chi_A$ and symmetric $\chi_S$ spin wave functions: 
\eq 
\chi_A \, = \, \sqrt{\frac{1}{2}} \ \biggl\{ 
\uparrow (\uparrow \downarrow - \downarrow \uparrow) \biggr\}\,, 
\quad\quad \chi_S \, = \, \sqrt{\frac{1}{6}}  \ \biggl\{ 
\uparrow (\uparrow \downarrow + \downarrow \uparrow)  
- 2 \downarrow \uparrow \uparrow \biggr\} \,. 
\en

\newpage

\begin{table} 
{\Large\bf List of Tables}

\vspace*{.5cm}

\begin{center}
{\bf Table 1.} 
Single charm $1/2^{+}$ baryons  

\vspace*{.5cm}

\def\arraystretch{1.2}
\begin{tabular}{|c|c|c|c|c|c|c|c|}
\hline 
\,\, Notation \,\,  & \,\, Content \,\, &  \,\, $J^P$ \,\, &
\,\, SU(3) \,\, & \,\, $I_3$ \,\, & \,\, S \,\,
& \,\, C \,\, & \,\, Mass (GeV)\,\, \\[2mm]
\hline
 $\Lambda_{c}^{+}$ & $c[ud]$ &  $1/2^+ $  &
 $\bar 3$ & 0 &
 0 & 1 & $2.286$ \\
 $\Xi_{c}^{+}$ & $c[su]$ & $1/2^+ $  &
 $\bar 3$  & 1/2 &
 -1 & 1 & $2.466$ \\
 $\Xi_{c}^{0}$ & $c[sd]$ & $1/2^+ $  &
 $\bar 3$  & -1/2 &
 -1 & 1 & $2.472$ \\
\hline
 $\Sigma_{c}^{++}$ & $cuu$ & $1/2^+ $  &
$6$ & 1 &
 0 & 1 & $2.453$ \\
 $\Sigma_{c}^{+}$ & $c\{ud\}$ & $1/2^+ $  &
$6$ & 0 &
 0 & 1 & $2.451$\\
 $\Sigma_{c}^{0}$ & $cdd$ & $1/2^+ $  &
$6$ & -1 &
 0 & 1 & $2.452$ \\
 $\Xi_{c}^{\prime +}$ & $c\{su\}$ & $1/2^+ $  &
$6$ & 1/2 &
 -1 & 1 & $2.574$ \\
 $\Xi_{c}^{\prime 0}$ & $c\{sd\}$ & $1/2^+ $  &
$6$ & -1/2 &
 -1 & 1 & $2.579$\\
 $\Omega_{c}^{0}$ & $css$ & $1/2^+ $  &
$6$ & 0 &  -2 & 1 & $2.698$ \\ 
\hline
\end{tabular}
\end{center}
\end{table}

\begin{table}
\begin{center}
{\bf Table 2.} 
Single bottom $1/2^{+}$ baryons  

\vspace*{.5cm}
\def\arraystretch{1.2}
\begin{tabular}{|c|c|c|c|c|c|c|c|}
\hline
\,\, Notation \,\,  & \,\, Content \,\, &  \,\, $J^P$ \,\, &
\,\, SU(3) \,\, & \,\, $I_3$ \,\, & \,\, S \,\,
& \,\, B \,\, & \,\, Mass (GeV)\,\, \\[2mm]
\hline
 $\Lambda_{b}$ & $b[ud]$ & $1/2^+$ & $\bar 3$ & 0 &
0 & 1 & $ 5.624$ \\
 $\Xi_{b}^{0}$ & $b[su]$ & $1/2^+$ & $\bar 3$ & 1/2  &
 -1 & 1 & $ 5.80 $ \\
 $\Xi_{b}^{-}$ & $b[sd]$ & $1/2^+$ & $\bar 3$ & -1/2 &
 -1 & 1 & $ 5.80 $ \\ \hline
 $\Sigma_{b}^{+}$ & $buu$ & $1/2^+$ & $6$ & 1 &
 0 & 1 & $ 5.82 $ \\
 $\Sigma_{b}^{0}$ & $b\{ud\}$ & $1/2^+$ & $6$ & 0 &
 0 & 1 & $ 5.82 $ \\
 $\Sigma_{b}^{-}$ & $bdd$ & $1/2^+$ & $6$ & -1 &
 0 & 1 & $ 5.82 $ \\
 $\Xi_{b}^{\prime 0}$ & $b\{su\}$ & $1/2^+$ & $6$ & 1/2 &
 -1 & 1 & $ 5.94 $ \\
 $\Xi_{b}^{\prime -}$ & $b\{sd\}$ & $1/2^+$ & $6$ & -1/2 &
 -1 & 1 & $ 5.94 $ \\
$\Omega_{b}^{-}$ & $bss$ & $1/2^+$ & $6$ & 0 &
 -2 & 1 & $ 6.04 $\\ 
\hline
\end{tabular}
\end{center}
\end{table}

\begin{table}
\begin{center}
{\bf Table 3.} 
Double and triple heavy $1/2^{+}$ baryons  

\vspace*{.5cm}
\def\arraystretch{1.1}
\begin{tabular}{|c|c|c|c|c|c|c|c|}
\hline 
\,\, Notation \,\,  & \,\, Content \,\, &  \,\, $J^P$ \,\, &
\,\, $I_3$ \,\, & \,\, S  & \,\, C \,\,
& \,\, B \,\, & \,\, Mass (GeV)\,\, \\[2mm]
\hline
$\Xi_{cc}^{++}$ & $u\{cc\}$ & $1/2^+$ & 1/2 &
 0  & 2 & 0 & $ 3.519$ \\
$\Xi_{cc}^{+}$ & $d\{cc\}$ & $1/2^+$ & -1/2 &
 0  & 2 & 0 & $ 3.519$ \\
$\Omega_{cc}^{+}$ & $s\{cc\}$ & $1/2^+$ & 0 &
 -1 & 2 & 0 & $ 3.59$ \\ \hline 
$\Xi_{bb}^{0}$ & $u\{bb\}$ & $1/2^+$ & 1/2 &
  0 & 0 & 2 & $10.09$ \\
$\Xi_{bb}^{-}$ & $d\{bb\}$ & $1/2^+$ & -1/2 &
  0 & 0 & 2 & $10.09$ \\
$\Omega_{bb}^{-}$ & $s\{bb\}$ & $1/2^+$ & 0 &
 -1 & 0 & 2 & $ 10.18$ \\ \hline
$\Xi_{cb}^+$ & $u[cb]$ & $1/2^+$ & 1/2 &
 0  & 1 & 1 & $ 6.82$ \\
$\Xi_{cb}^0$ & $d[cb]$ & $1/2^+$ & -1/2 &
 0  & 1 & 1 & $ 6.82$ \\
$\Omega_{cb}^{0}$ & $s[cb]$ & $1/2^+$ & 0 &
 -1  & 1 & 1 & $ 6.91$ \\ \hline
$\Xi_{cb}^{\prime \, +}$ & $u\{cb\}$ & $1/2^+$ & 1/2 &
 0  & 1 & 1 & $ 6.85$ \\
$\Xi_{cb}^{\prime \, 0}$ & $d\{cb\}$ & $1/2^+$ & -1/2 &
 0  & 1 & 1 & $ 6.85$ \\
$\Omega_{cb}^{\prime 0}$ & $s\{cb\}$ & $1/2^+$ & 0 &
 -1  & 1 & 1 & $ 6.93$ \\ \hline 
$\Omega_{ccb}^{+}$ & $bcc$ & $1/2^+$ & 0 & 0 & 2 & 1 & $8.0$ \\
$\Omega_{cbb}^{0}$ & $cbb$ & $1/2^+$ & 0 & 0 & 1 & 2 & $11.5$ \\
\hline
\end{tabular}
\end{center}
\end{table}

\begin{table}
\begin{center}
{\bf Table 4.} 
Magnetic moments of single heavy baryons 
(in units of $\mu_{N}$) 

\vspace*{.5cm}
\def\arraystretch{2.5}
\begin{tabular}{|c|c|c|c|c|}
\hline
Baryon & \multicolumn{3}{|c|}{RQM} & NRQM \\
\cline{2-4} & full & HQL BCF & HQL BCF+HQP & \\
\hline
$\Lambda_{c}^{+}$  & 0.42 (0.41; 0.01)  
                   & 0.38 (0.38; 0.003) 
                   & 0.37 (0.37; 0) 
                   & 0.37 (0.37; 0) \\
$\Lambda_{b}^{0}$  & -0.06 (-0.06; 0.002) 
                   & -0.06 (-0.06; 0.001) 
                   & -0.06 (-0.06; 0) 
                   & -0.06 (-0.06; 0) \\
$\Xi_{c}^{+}$      & 0.41 (0.40; 0.01) 
                   & 0.37 (0.37; 0.01) 
                   & 0.37 (0.37, 0) 
                   & 0.37 (0.37; 0) \\
$\Xi_{c}^{0}$      & 0.39 (0.40; -0.01) 
                   & 0.37 (0.37; -0.004) 
                   & 0.37 (0.37; 0) 
                   & 0.37 (0.37; 0) \\
$\Xi_{b}^{0}$      & -0.06 (-0.06; 0.002) 
                   & -0.06 (-0.06; 0.001) 
                   & -0.06 (-0.06; 0) 
                   & -0.06 (-0.06; 0) \\
$\Xi_{b}^{-}$      & -0.06 (-0.06; -0.003) 
                   & -0.06 (-0.06: -0.001)
                   & -0.06 (-0.06; 0) 
                   & -0.06 (-0.06; 0) \\
\hline 
$\Xi_{c}^{\prime +}$ & 0.47 (-0.11; 0.58) 
                     & 0.10 (-0.11; 0.21)
                     & 0.08 (-0.12; 0.20) 
                     & 0.51 (-0.12; 0.63) \\ 
$\Xi_{c}^{\prime 0}$ & -0.95 (-0.11; -0.84)  
                     & -0.38 (-0.11; -0.27)
                     & -0.37 (-0.12; -0.25) 
                     & -0.98 (-0.12; -0.86) \\ 
$\Xi_{b}^{\prime 0}$ & 0.66 (0.02; 0.64) 
                     & 0.22 (0,02; 0.20) 
                     & 0.22 (0.02; 0.20) 
                     & 0.65 (0.02; 0.63) \\
$\Xi_{b}^{\prime -}$ & -0.91 (0.02; -0.93) 
                     & -0.23 (0.02; -0.25) 
                     & -0.23 (0.02; -0.25) 
                     & -0.84 (0.02; -0.86) \\
\hline
$\Sigma_{c}^{++} $   & 1.76 (-0.11; 1.87) 
                     & 0.58 (-0.11; 0.69) 
                     & 0.53 (-0.12; 0.65) 
                     & 1.86 (-0.12; 1.98) \\
$\Sigma_{c}^{+}$     & 0.36 (-0.11; 0.47) 
                     & 0.06 (-0.11; 0.17) 
                     & 0.04 (-0.12; 0.16) 
                     & 0.37 (-0.12; 0.49) \\
$\Sigma_{c}^{0}$     & -1.04 (-0.11; -0.93) 
                     & -0.46 (-0.11; -0.35) 
                     & -0.44 (-0.12; -0.32) 
                     & -1.11 (-0.12; -0.99) \\
$\Sigma_{b}^{+}$     & 2.07 (0.02; 2.05) 
                     & 0.68 (0.02; 0.66) 
                     & 0.67 (0.02; 0.65) 
                     & 2.01 (0.02; 1.99) \\
$\Sigma_{b}^{0}$     & 0.53 (0.02; 0.51) 
                     & 0.18 (0.02; 0.16) 
                     & 0.18 (0.02; 0.16) 
                     & 0.52 (0.02; 0.50) \\
$\Sigma_{b}^{-}$     & -1.01 (0.02; -1.03) 
                     & -0.31 (0.02; -0.33) 
                     & -0.30 (0.02; -0.32) 
                     & -0.97 (0.02; -0.99) \\
\hline
$\Omega_{c}^{0}$     & -0.85 (-0.11; -0.74) 
                     & -0.32 (-0.11; -0.21) 
                     & -0.31 (-0.12; -0.19) 
                     & -0.85 (-0.12; -0.73) \\
$\Omega_{b}^{-}$     & -0.82 (0.02; -0.84) 
                     & -0.17 (0.02; -0.19) 
                     & -0.17 (0.02; -0.19)
                     & -0.71 (0.02; -0.73) \\
\hline
\end{tabular}
\end{center}
\end{table}

\begin{table}
\begin{center}
{\bf Table 5.} 
Predictions for the HHChPT coupling constant 
$c_{S}$   

\vspace*{.5cm}
\def\arraystretch{2.5}
\begin{tabular}{|c|c|c|c|c|}
\hline
Baryon & $\Sigma_{Q\{qq^\prime\}}$ 
       & $\Xi_{Q\{us\}}$ 
       & $\Xi_{Q\{ds\}}$ 
       & $\Omega_{Q\{ss\}}$ \\ 
\hline
$c_S$  & 0.45 & 0.55 & 0.35 & 0.26 \\
\hline

\hline
\end{tabular}
\end{center}
\end{table}

\newpage 

\begin{table}
\begin{center}
{\bf Table 6.} 
Magnetic moments of double and triple heavy baryons 
(in units of $\mu_{N}$) 

\vspace*{.5cm}
\def\arraystretch{2.5}
\begin{tabular}{|c|c|c|c|}
\hline
Baryon & \multicolumn{2}{|c|}{RQM} & NRQM \\
\cline{2-3} & full & HQL BCF & \\
\hline
$\Xi_{cc}^{++}$  & 0.13 (0.52; -0.38) 
                 & 0.25 (0.51; -0.26) 
                 & -0.01 (0.49; -0.50) \\
$\Xi_{cc}^{+}$   & 0.72 (0.52; 0.20) 
                 & 0.64 (0.51; 0.13) 
                 & 0.74 (0.49; 0.25) \\
$\Xi_{bb}^{0}$   & -0.53 (-0.06; -0.47)
                 & -0.42 (-0.08; -0.34)
                 & -0.58 (-0.08; -0.50) \\
$\Xi_{bb}^{-}$   & 0.18 (-0.06; 0.24)
                 & 0.09 (-0.08; 0.17)    
                 & 0.17 (-0.08; 0.25) \\
$\Omega_{cc}^{+}$ & 0.67 (0.53; 0.14)
                  & 0.60 (0.50; 0.10)
                  & 0.67 (0.49; 0.18) \\
$\Omega_{bb}^{-}$ & 0.04 (-0.08; 0.12)
                  & 0.14 (-0.06; 0.20)
                  & 0.10 (-0.08; 0.18) \\
\hline
$\Xi_{cb}^{+}$ & 1.52 (0.002; 1.52)  
               & 0.75 (0.001; 0.75)       
               & 1.49 (0; 1.49) \\
$\Xi_{cb}^{0}$ & -0.76 (0.002, -0.76)  
               & -0.38 (0.001; -0.38)       
               & -0.74 (0; -0.74) \\
$\Xi_{cb}^{\prime +}$ & -0.12 (0.24; -0.36)
                      & 0.18 (0.42; -0.24)
                      & -0.29 (0.21; -0.50) \\
$\Xi_{cb}^{\prime 0}$ & 0.42 (0.24; 0.18) 
                      & 0.54 (0.42; 0.12)   
                      & 0.46 (0.21; 0.25) \\
$\Omega_{cb}^{0}$     & -0.61 (0.002; -0.61) 
                      & -0.26 (0.001; -0.26)
                      & -0.55 (0; -0.55) \\ 
$\Omega_{cb}^{\prime 0}$ & 0.45 (0.25; 0.20) 
                         & 0.50 (0.42; 0.08) 
                         & 0.39 (0.21; 0.18) \\ 
\hline
$\Omega_{ccb}^{+}$       & 0.53 (0.02; 0.51)
                         & 0.14 (0.02; 0.12)    
                         & 0.51 (0.02; 0.49) \\ 
$\Omega_{cbb}^{0}$       & -0.20 (-0.08; -0.12)
                         & -0.13 (-0.05; -0.08)
                         & -0.20 (-0.08; -0.12) \\ 
\hline
\end{tabular}
\end{center}
\end{table}

\newpage 

\begin{table}
\begin{center} 
{\bf Table 7.} 
Heavy baryon wave functions and magnetic moments \\ 
in the nonrelativistic quark model, where $q,q^\prime = u$ or $d$ and 
$Q,Q^\prime = c$ or $b$.  

\vspace*{.5cm}
\def\arraystretch{2.5}
\begin{tabular}{|c|c|c|} 
\hline
  $\;\;$ Baryon $\;\;$ & $\;\;\;\;\;\;\;$ Wave function $\;\;\;\;\;\;\;$ 
& $\;\;\;\;\;$ Magnetic moment $\;\;\;\;\;$ \\ 
\hline 
$\Lambda_{Q[ud]}$ &
$\frac{1}{\sqrt{2}} \, Q ( ud - du ) \,\,\, \chi_A$ & 
$\frac{\displaystyle e_Q}{\displaystyle 2m_Q}$ \\ 
\hline
$\Xi_{Q[qs]}$   & 
$\frac{1}{\sqrt{2}} \, Q ( qs - sq ) \,\,\, \chi_A$ & 
$\frac{\displaystyle e_Q}{\displaystyle 2m_Q}$ \\
\hline 
$\Sigma_{Q\{qq^\prime\}}$ & 
$\frac{1}{\sqrt{2}} \, 
Q ( qq^\prime + q^\prime q ) \,\,\, \chi_S$ & 
$- \frac{\displaystyle e_Q}{\displaystyle 6m_Q} 
 + \frac{\displaystyle e_q}{\displaystyle 3m_q} 
+ \frac{\displaystyle e_{q^\prime}}{\displaystyle 3m_{q^\prime}}$ \\ 
\hline 
$\Omega_{Q\{ss\}}$ & 
$Q ss \,\,\, \chi_S$ & 
$- \frac{\displaystyle e_Q}{\displaystyle 6m_Q} 
 + \frac{\displaystyle 2e_s}{\displaystyle 3m_s}$ \\ 
\hline 
$\Xi_{q\{QQ^\prime\}}$ & 
$\frac{1}{\sqrt{2}} q ( Q Q^\prime + Q^\prime Q) \,\,\, \chi_S$ & 
$- \frac{\displaystyle e_q}{\displaystyle 6m_q} 
 + \frac{\displaystyle e_Q}{\displaystyle 3m_Q} 
 + \frac{\displaystyle e_Q^\prime}{\displaystyle 3m_Q^\prime} 
$ \\ 
\hline 
$\Omega_{s\{QQ\}}$ & $s QQ \,\,\, \chi_S$ & 
$- \frac{\displaystyle e_s}{\displaystyle 6m_s} 
 + \frac{\displaystyle 2e_Q}{\displaystyle 3m_Q}$ \\ 
\hline  
$\Xi_{q[cb]}$ & 
$\frac{1}{\sqrt{2}} \, q (cb - bc) \,\,\, \chi_A$ & 
$\frac{\displaystyle e_q}{\displaystyle 2m_q}$ \\ 
\hline 
$\Xi_{q\{cb\}}$ & 
$\frac{1}{\sqrt{2}} \, q (cb + bc) \,\,\, \chi_S$ & 
$- \frac{\displaystyle e_q}{\displaystyle 6m_q} 
 + \frac{\displaystyle e_c}{\displaystyle 3m_c} 
 + \frac{\displaystyle e_b}{\displaystyle 3m_b}$ \\ 
\hline
$\Omega_{s[cb]}$ 
& $\frac{1}{\sqrt{2}} \, s (cb - bc) \,\,\, \chi_A$ &  
$\frac{\displaystyle e_s}{\displaystyle 2m_s}$ \\ 
\hline 
$\Omega_{s\{cb\}}$ & 
$\frac{1}{\sqrt{2}} \, s (cb + bc) \,\,\, \chi_S$ & 
$- \frac{\displaystyle e_s}{\displaystyle 6m_s} 
 + \frac{\displaystyle e_c}{\displaystyle 3m_c} 
 + \frac{\displaystyle e_b}{\displaystyle 3m_b}$ \\ 
\hline 
$\Omega_{b\{cc\}}$ & $b cc \,\,\, \chi_S$ &  
$- \frac{\displaystyle e_b}{\displaystyle 6m_b} 
 + \frac{\displaystyle 2e_c}{\displaystyle 3m_c}$ \\ 
\hline 
$\Omega_{c\{bb\}}$ & $c bb \,\,\, \chi_S$ & 
$- \frac{\displaystyle e_c}{\displaystyle 6m_c} 
 + \frac{\displaystyle 2e_b}{\displaystyle 3m_b}$ \\ 
\hline
\end{tabular}
\end{center} 
\end{table}

\newpage 

\begin{figure}
\begin{center}

\vspace*{.75cm}
\epsfig{file=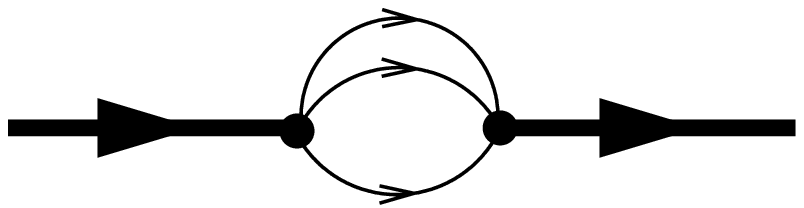,height=1.4cm} 

\vspace*{1cm}
\centerline{Fig.1 Baryon mass operator} 

\vspace*{2.25cm}

\epsfig{file=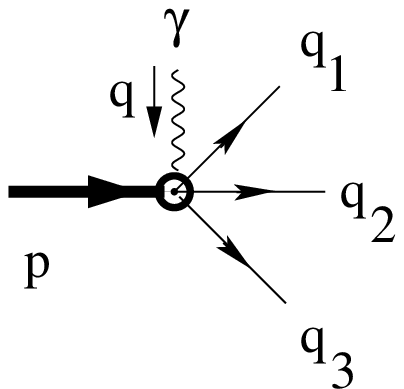,height=3.4cm}

\vspace*{1cm} 
\centerline{Fig.2 Coupling vertex of baryon, photon and three quarks} 
\label{fig1}

\vspace*{2.25cm}

\epsfig{file=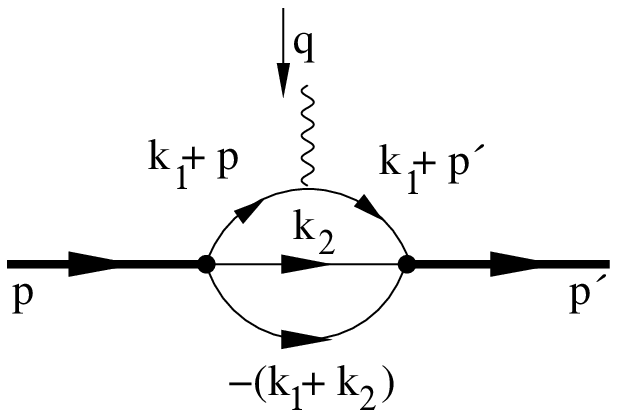, height=4.2cm}
\centerline{(a)} 
\epsfig{file=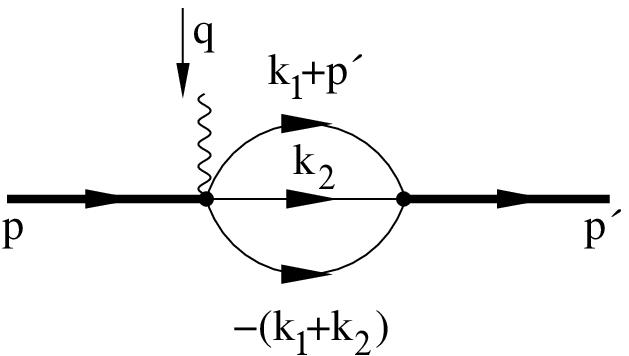, height=3.0cm}\hspace{1cm}
\epsfig{file=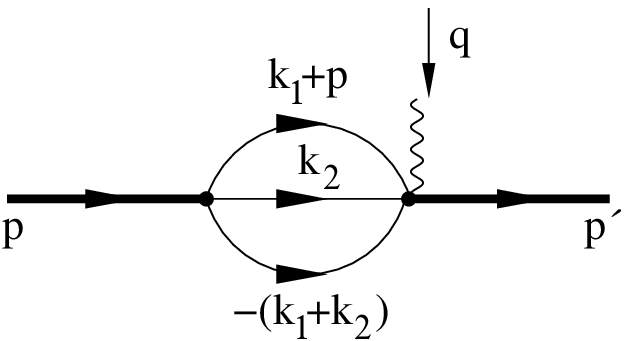, height=3.0cm}

\vspace*{.5cm}

\centerline{\hspace*{1cm} (b) \hspace*{6cm}(c) \hspace*{1cm} } 

\vspace*{1cm}

\centerline{Fig.3 Diagrams contributing to the baryon electromagnetic 
vertex function: triangle (a), bubble (b) and (c) diagrams.} 
\label{fig8}
\end{center}
\end{figure}

\end{document}